\documentclass[fleqn,usenatbib]{mnras}

\usepackage{newtxtext,newtxmath}
\usepackage[T1]{fontenc}
\DeclareRobustCommand{\VAN}[3]{#2}
\let\VANthebibliography\thebibliography
\def\thebibliography{\DeclareRobustCommand{\VAN}[3]{##3}\VANthebibliography}

\usepackage{graphicx}	% Including figure files
\usepackage{amsmath}	% Advanced maths commands

\usepackage{amssymb}	% Extra maths symbols

\def\xmm{{\it XMM-Newton}}
\def\nustar{{\it NuSTAR}}
\def\athena{{\it ATHENA}}
\def\xrism{{\it XRISM}}
\newcommand{\yrxu}[1]{\textcolor{black}{#1}}
\title[Ejection-accretion connection in 1H 1934-063]{Ejection-accretion connection in NLS1 AGN 1H 1934-063}

\author[Y. Xu et al.]{Y. Xu$^{1,2}$,\thanks{E-mail: yerong.xu@inaf.it}
    C. Pinto$^{1}$,
    E. Kara$^{3}$,
    M. Masterson$^{3}$,
    J. A. Garc\'ia$^{4,5}$,
    A. C. Fabian$^{6}$,
    M. L. Parker$^{6}$,
    \newauthor  
    D. Barret$^{7}$,
    W. N. Alston$^{8}$,
    G. Cusumano$^{1}$
\\
$^{1}$INAF - IASF Palermo, Via U. La Malfa 153, I-90146 Palermo, Italy\\
$^{2}$Universit\`a degli Studi di Palermo, Dipartimento di Fisica e Chimica, via Archirafi 36, I-90123 Palermo, Italy\\
$^{3}$MIT Kavli Institute for Astrophysics and Space Research, Cambridge, MA 02139, USA\\
$^{4}$Cahill Center for Astronomy and Astrophysics, California Institute of Technology, Pasadena, CA 91125, USA\\
$^{5}$Dr. Karl Remeis-Observatory and Erlangen Centre for Astroparticle Physics, Sternwartstr.~7, 96049 Bamberg, Germany\\
$^{6}$Institute of Astronomy, Madingley Road, CB3 0HA Cambridge, United Kingdom\\
$^{7}$Universit\'e de Toulouse, CNRS, IRAP, 9 Avenue du colonel Roche, BP 44346, 31028 Toulouse Cedex 4, France\\
$^{8}$European Space Agency (ESA), European Space Astronomy Centre (ESAC), E-28691 Villanueva de la Canada, Madrid, Spain\\
}

\date{Accepted XXX. Received YYY; in original form ZZZ}

\pubyear{2021}

\begin{document}
\label{firstpage}
\pagerange{\pageref{firstpage}--\pageref{lastpage}}
\maketitle

\begin{abstract}
Accretion and ejection of matter in active galactic nuclei (AGN) are tightly connected phenomena and represent fundamental mechanisms regulating the growth of the central supermassive black hole and the evolution of the host galaxy. However, the exact physical processes involved are not yet fully understood. We present a high-resolution spectral analysis of a simultaneous \xmm\ and \nustar\ observation of the narrow line Seyfert 1 (NLS1) AGN 1H 1934-063, during which the X-ray flux dropped by a factor of $\sim6$ and subsequently recovered within 140 kiloseconds. By means of the time-resolved and flux-resolved X-ray spectroscopy, we discover a potentially variable warm absorber and a relatively stable ultra-fast outflow (UFO, $v_\mathrm{UFO}\sim-0.075\,c$) with a mild ionization state ($\log(\xi/\mathrm{erg\,cm\,s^{-1})}\sim1.6$). The detected emission lines (especially a strong and broad feature around 1\,keV) are of unknown origin and cannot be explained with emission from plasmas in photo- or collisional-ionization equilibrium. Such emission lines could be well described by a strongly blueshifted ($z\sim-0.3$) secondary reflection off the base of the equatorial outflows, which may reveal the link between the reprocessing of the inner accretion flow photons and the ejection. However, this scenario although being very promising is only tentative and will be tested with future observations.

\end{abstract}


%%%%%%%%%%%%%%%%%%%%%%%%%%%%%%%%%%%%%%%%%%%%%%%%%%

%%%%%%%%%%%%%%%%% BODY OF PAPER %%%%%%%%%%%%%%%%%%

\section{Introduction}
An active galactic nucleus (AGN) is a compact region at the center of a galaxy, powered by accretion of matter onto a central supermassive black hole (SMBH), producing electromagnetic radiation from radio to gamma-rays. In particular, the emission in the X-ray band is a probe for the very innermost region of the accretion disk, as it mainly originates from the  corona around the SMBH, where the ultra violet (UV) or optical photons from the surrounding accretion disk are being inverse-Compton scattered to produce a power-law continuum in the X-rays \citep[e.g.][]{1980Sunyaev,1993Haardt}. The reprocessing of the inner coronal photons reflected off the disk could create a series of fluorescent lines and a Compton hump ($>10\,$keV) in the spectrum. There is often a soft excess below 2\,keV, whose nature is still under debate with two main scenarios involving a warm Comptonization component \citep[e.g.][]{2012Done} or the relativistically blurred reflection \citep[e.g.][]{2019Garc} . 

Besides accreting a large amount of gas, AGN can release part of the accumulated energy via the ejection of powerful outflows within the disk  \citep[for a review see][]{2012Fabian}. If the output energy is above $0.5\mbox{--}5\%$ of the Eddington luminosity of the AGN, it can have a profound impact onto the evolution of the host galaxy \citep[e.g.][]{2005DiMatteo,2010Hopkins}. Outflows can expel or heat the surrounding interstellar medium, thus affecting the star formation and further accretion of matter onto the SMBH \citep{2012Zubovas,2017Maiolino}. Therefore, studying outflows responding to the accretion flow is essential to understand the physical link between the AGN and its host galaxy.

UFOs are the most extreme winds launched by AGN due to their mildly relativistic speeds greater than 10000\,km/s and originate from the inner region of the accretion disk \citep[e.g.][]{2002Chartas,2010Tombesi}. They were initially discovered through observations of blueshifted absorption features from Fe K-shell transitions of AGN (e.g. APM 08279+5255, \citealt{2002Chartas}, Mrk 766, \citealt{2003Pounds}). UFOs are expected to carry sufficient kinetic energy to affect the host galaxy since their output energy is often above $0.5\mbox{--}5\%$ of the Eddington luminosity \citep[e.g.][]{2013Tombesi,2015Nardini}. \yrxu{Their origin and connection with the most common slow winds (i.e. warm absorbers, $v\lesssim5000$ km/s) are still not well understood \citep{2021Laha}.} They may be driven either by the radiation pressure \citep[e.g.][]{2000Proga,2010Sim,2016Hagino} or by magneto-rotational forces \citep[MHD models, e.g.][]{2010Fukumura,2015Fukumura} or a combination of both. NLS1 AGN host low-mass, high-accretion-rate SMBHs with a strong radiation field, which is therefore expected to be a dominant driving force for UFOs \citep[see review by][]{2007Komossa}. Previous works have shown that UFOs can have some degree of variability and that their properties seem to be affected by variations in the source X-ray luminosity, at least in three sources for which a large amount of data was available: PDS 456 \citep{2017Matzeu}, IRAS 13224-3809 \citep{2017Parker,2018Pinto} and 1H 0707-495 \citep{2021Xu}. The wind structure determined by the acceleration mechanism, has also been suggested to play a role for the UFO variability recently \citep[e.g. MCG-03-58-007,][]{2021Braito,2021Mizumoto}. It is therefore of interest to investigate the variability of the outflows in other NLS1s.

Furthermore, by assuming that the escape velocity is equal to that observed in the outflowing gas, the fastest UFOs are thought to be launched in the inner region of the accretion disk, which is the same region that is probed by the inner-disk reflection spectroscopy. If the inner layer of the outflow is thick enough, which is possible at or above the Eddington limit, it may reflect part of X-ray disk/corona emission producing a Doppler shifted reflection signal. This scenario has been proposed to explain the strongly blueshifted Fe K emission line in a super-Eddington AGN Swift J1644+57 \citep{2016Kara}. Therefore the reflection emission might reveal the physical properties of the wind launching region and improve our understanding of UFOs.

1H 1934-063 (hereafter 1H 1934; also known as IGR J19378-0617, SS 442 and IRAS 19348-0619) is a radio-quiet NLS1 \citep[][]{2011Panessa} galaxy at a redshift of $z=0.0102$ \citep{2008Rodriguez}, which ranked seventh in 10$\mbox{--}$20 ks excess variance among 161 AGN in the \xmm\ archive \citep[CAIXA;][]{2012Ponti}. It hosts a central supermassive black hole (SMBH) with a mass of $M_\mathrm{BH}=3\times10^6\,M_\odot$, estimated from FWHM H$\beta$ \citep{2000Rodriguez-Ardila,2008Malizia}. 
%1H 1934 was reported by two {\it INTEGRAL}/IBIS catalogs \citep{2004Molkov,2007Bird}, the MAXI 37-month catalog \citep{2013Hiroi}, the Swift BAT 70-month catalog \citep{2013Baumgartner} and the Suzaku catalog \citep{2020Waddell}. 
This source was twice observed by \xmm\ in 2009 \citep[18\,ks,][]{2011Panessa} and 2015 \citep[140\,ks,][]{2018Frederick} joint with a 65\,ks \nustar\ observation. 1H 1934 has two different measurements of the black hole spin, $a_\star<0.1$ \citep{2018Frederick} and $a_\star>0.4$ \citep{2019Jiang}, depending on whether a high density was adopted in the reflection model. \citet{2018Frederick} also discovered a time lag ($\sim$20\,s) of the disk reflection components behind the coronal power-law continuum, indicating a $9\pm4\,R_\mathrm{g}$ ($R_\mathrm{g}\equiv GM_\mathrm{BH}/c^2$) distance between the corona and the accretion disk.

The fact that this AGN is bright ($F_\mathrm{X}\sim5\times10^{-11}\,\mathrm{erg/cm^{2}/s}$) and rapidly variable in the X-ray energy band may enable us to study any potential outflow and its variability. We present the soft X-ray analysis based on the concurrent \xmm\ and \nustar\ observation of 1H 1934 (PI: E. Kara). This paper is organized as follows. We present the data reduction procedure and products in Section~\ref{sec:data&products}. Details on our analysis and results are shown in Section~\ref{sec:results}. We discuss the results and provides our conclusions in Section~\ref{sec:discussion} and Section~\ref{sec:conclusion}, respectively.

\section{Data reduction and products}\label{sec:data&products}
\subsection{Data reduction}\label{subsec:reduction}
1H 1934 was simultaneously observed by \xmm\ \citep[Obs. ID: 0761870201;][]{2001Jansen} and \nustar\ \citep[Obs. ID: 60101003002;][]{2013Harrison} on 2015 October 1-3 with gross exposure time of 140 and 65\,ks, respectively. \xmm\ consists of the European Photon Imaging Camera (EPIC) including two EPIC-MOS CCDs \citep{2001Turner} and an EPIC-pn CCD \citep{2001Struder}, the Reflection Grating Spectrometers \citep[RGS;][]{2001denHerder}, and the Optical Monitor \citep[OM;][]{2001Mason}. We utilize the data from the EPIC and two Focal Plane Module (FPMA/B) onboard \nustar\ to determine the X-ray broadband spectrum. The results of this paper are mostly based on the high-resolution RGS spectrum. The OM spectrum is also included to get the Spectral Energy Distribution (SED) for photoionization modelling. 

The \xmm\ data are reduced with the Science Analysis System (SAS v19.1.0) and calibration files available on May 2021, following the standard SAS threads. In brief, the EPIC-pn and EPIC-MOS data are processed with {\small epproc} and {\small emproc} package, respectively. The filtering criteria of the background flare contamination are set at the standard values of 0.5 and 0.35 counts/sec (in the $10\mbox{--}12$\,keV) for pn and MOS separately. We extracted EPIC source spectra from a circular region of radius 30 arcsec, and background spectra from a nearby source-free region of the same radius. We do not find significant pile-up in both pn and MOS above 7\,keV, which was marginally reported in \citet{2018Frederick}. We stack MOS1 and MOS2 spectra to maximize the signal-to-noise. The EPIC spectra are grouped to over-sample the instrumental resolution by a factor of 3 and to a minimum of 30 counts per energy bin. The RGS data are processed with the {\small rgsproc} tool, for which background flares are excluded by a threshold of 0.2 counts/sec. We extract the first-order RGS spectra in a cross-dispersion region of 1' width and the background spectra by selecting photons beyond the 98\% of the source point-spread-function as default. We only use the good time intervals common to both RGS 1 and 2 and stack their spectra for the high signal-to-noise. We regroup the RGS spectrum so that each bin is not narrower than 1/3 of the spectral resolution. The OM spectra are reduced using the {\small omichain} pipeline, including all necessary calibration processes. We retrieve the canned response file from the ESA webpage\footnote{https://www.cosmos.esa.int/web/xmm-newton/om-response-files} for the UVW2 filter (2120\,\AA), which was only adopted during the observation.

\nustar\ observed 1H 1934 simultaneously with \xmm\ for exposure times of roughly 65\,ks per instrument. The reduction of the \nustar\ data is conducted following the standard procedures with the NuSTAR Data Analysis Software (NUSTARDAS v.2.0.0) and the updated calibration files from NuSTAR CALDB v20210427 using the {\small nupipeline} task. The source spectrum is extracted from a circular region of radius 80 arcsec and the background spectrum from a free-source region with a radius of 120 arcsec. The spectra are grouped to at least 30 counts per bin in order to have a sufficiently high signal-to noise ratio.

\subsection{Light curve}\label{subsec:lc}
\begin{figure}
	\includegraphics[width=\columnwidth, trim={90 120 20 70}]{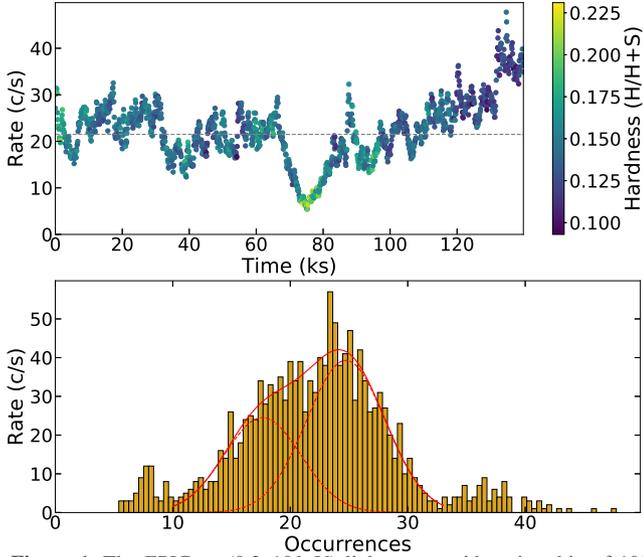}
    \caption{The EPIC-pn (0.3$\mbox{--}10$\,keV) light curve with a time bin of 100\,s ({\it \yrxu{top}}) and the corresponding count rate histogram ({\it \yrxu{bottom}}). The light curve is coded according to the hardness ratio. S and H denote the counts in the soft and hard energy bands defined as $0.3\mbox{--}2$ and $2\mbox{--}10$\,keV, respectively. The horizontal dash grey line indicates the threshold of the flux intervals, which is determined by the intersection (21.5 c/s) between the two Gaussian lines fitted (red lines in the \yrxu{bottom} panel) to the count rate histogram. The vertical grey region indicates the duration of flux dip, of which spectrum is extracted (see Section~\ref{subsec:system}).}
    \label{fig:lc_histogram}
\end{figure}

We show the EPIC-pn (0.3$\mbox{--}$10\,keV) light curve with a time bin of 100\,s and the corresponding count rate histogram in Fig.\ref{fig:lc_histogram}. The light curve is color-coded according to the hardness ratio (HR=H/H+S, H: $2\mbox{--}10$\,keV; S: $0.3\mbox{--}2$\,keV). 1H 1934 presents a softer-when-brighter behavior during the observation with a strong flux dip by a factor of $\sim6$, which was proposed to result from the strong light-bending effect as the change in accretion rate was excluded due to the stable UV flux \citep{2018Frederick}. To resolve any narrow and variable spectral features, we divide the entire light curve into two flux intervals, marked by the horizontal grey dash line, based on the intersection (21.5 c/s) between two Gaussians that were fitted to the count rate distribution (see red lines in the right panel of Fig.\ref{fig:lc_histogram}). We caution that this flux cut is just a way to carve up the data, not indicating two processes at different flux \citep{2019Alston}. Accordingly, the low- and high-flux spectra are extracted as well as the flux-resolved \nustar\ spectra at the exactly same time intervals. The ratio of total counts between the low- and high- flux is about 0.5. In addition, we also test the dip-resolved spectroscopy in section~\ref{subsec:system} (see Fig.\ref{fig:dip}) by extracting the dip spectrum, as marked by the grey region in Fig.\ref{fig:lc_histogram}.

%%%%%%%%%%%%%%%%%%%%%%%%%%%%%%%%%%%%%%%%%%%%%%%%%%
%%%%%%%%%%%%%%%%%%%%%%%%%%%%%%%%%%%%%%%%%%%%%%%%%%

\section{Results}\label{sec:results}
\subsection{Spectral Variability}\label{subsec:PCA+Fvar}
As we are mainly interested in the search for resolved spectral lines and evidence for outflows in 1H 1934, we start with a model-independent spectral-variability analysis. According to \citet{2015Parker,2017Parkerb,2021Parker}, the principal component analysis (PCA) and the fractional root-mean-square (RMS) variability amplitude ($F_\mathrm{var}$) spectra could identify a series of variability peaks in both the first PCA component and $F_\mathrm{var}$ spectrum corresponding to the strongest absorption lines from the UFO, because the UFO can be highly variable on timescale of hours or less and exhibit a rapid response to changes in the continuum. The PCA method performs a singular value decomposition (SVD) to decompose a matrix of spectra, and split it according to the given time bin, into a set of orthogonal PCs which account for the majority of the coherent variability of the source. The $F_\mathrm{var}$ spectrum is used to calculate the total RMS normalized excess variance above the expected noise level as a function of energy.

\begin{figure}
	\includegraphics[width=\columnwidth, trim={30 80 50 0}]{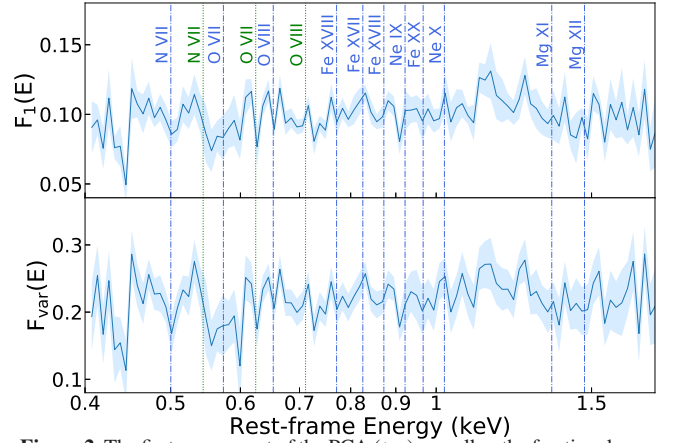}
    \caption{The first component of the PCA (\yrxu{{\it top}}) as well as the fractional excess variability (\yrxu{{\it bottom}}) of RGS spectrum are shown. The rest-frame positions of the ion transition lines among RGS band are marked through the \yrxu{vertical {\it blue dashdot} lines.} Three highly blueshifted ($z\sim-0.078$) absorption features are also indicated with \yrxu{vertical {\it green dotted} lines. The line markers are the same for following figures.}} 
    %The single trial significance obtained from the Gaussian line scan with different line width (100 to 10000\,km/s) over the RGS spectrum is displayed in the {\it middle}. The rest-frame positions of the ion transition lines among RGS band are marked through the dash {\it blue} lines. Three highly blueshifted ($z\sim-0.078$) absorption features are also indicated with {\it green} lines. The {\it grey} region marks the significance 3$\sigma$. The true significance of the residuals ({\it lower}) is obtained from the Monte Carlo simulations. The {\it grey} region represents 99\% significance level and different line widths are marked by different colors.}
    \label{fig:PCA}
\end{figure}

We apply the PCA method described in \citet{2015Parker}, using the code of \citet{2017Parkerb}, to the RGS data in order to search for narrow spectral features. We adopt 10\,ks time bins and 100 energy points in the logarithmic space between 0.3 and 10 keV. The light curve of \xmm\ is accordingly split into 13 segments. The first principal component (PC1) of RGS is shown in the \yrxu{top} panel of Fig.\ref{fig:PCA} for clarity. The RGS PC1 spectrum shows some dips in correspondence to the energies of the dominant X-ray transitions. Later on we will attempt their identification through spectral modelling. However, there are some peaks that do not always match such transitions, possibly indicating the presence of Doppler shifted variable lines. 
%%%%%%%%%%%%%%%%%%%%%%%%%%%%%%%%%%%%%%%%%%%%%%%%%%

Following the method introduced by \citet{2003Vaughan}, we adopt the same spectra used for the PCA to compute the RMS spectrum. The $F_\mathrm{var}$ is defined as $F_\mathrm{var}=\sqrt{\frac{S^2-\overline{\sigma^2_\mathrm{err}}}{\bar{x}^2}}$, where $\bar{x}$ is the mean flux of the time segments at given energy bin, $\overline{\sigma_\mathrm{err}^{2}}=\frac{1}{N}\sum\limits_{i=1}^{N}\sigma^2_\mathrm{err,i}$ is the mean square error, $N$ is the number of time segments, and $S^2=\frac{1}{N}\sum\limits_{i=1}^{N}(x_i-\bar{x})^2$ is the mean square variance. The result is displayed in the \yrxu{bottom} panel of Fig.\ref{fig:PCA}. The shape of the RMS spectrum is remarkably similar to that of PCA, which only shows the amplitude of the {\it correlated} variability.

\subsection{XMM-NuSTAR Continuum Modelling}\label{subsec:continuum}
We begin the broadband X-ray spectroscopy by fitting the time-averaged RGS, EPIC, FPM spectra simultaneously using the XSPEC (v12.11.1) package \citep{1996Arnaud}. The instrumental differences are taken into account by employing a variable cross-calibration factor \citep[except for RGS fixed to unity;][]{2015Madsen}. We adopt the $\chi^2$ statistics and estimate all parameter uncertainties at the 90\% confidence level corresponding to $\Delta\chi^2=2.71$ in this paper. The luminosity calculations are based on the assumptions of $H_0=70\,$km/s/Mpc, $\Omega_\Lambda=0.73$ and $\Omega_M=0.27$. After checking the consistency of residuals between EPIC and RGS spectra to the continuum model in soft energies, we use the RGS data between 0.4$\mbox{--}$1.77\,keV and EPIC (pn+MOS) data between 1.77$\mbox{--}$10\,keV for spectral fittings, because the relatively low resolution but higher count rate of EPIC may effect the detection and identification of atomic spectral features by increasing model degeneracies. Due to the background contamination, we limit the analysis of the \nustar\ (FPMA/B) spectra to the $3\mbox{--}30$\,keV energy band. The same selection applies to the flux-resolved spectra.

We adopt a similar model to the best-fit one found in \citet{2018Frederick}: {\tt tbabs*zashift*(diskbb+relxilllpCp)}, to explain the broadband continuum. Briefly, this model takes into account the galactic hydrogen absorption ({\tt tbabs}) with the solar abundance calculated by \citet{2000Wilms}, the redshift of the source ({\tt zashift}), the X-ray soft excess in the form of a multi-colour disk blackbody ({\tt diskbb}), and the coronal continuum as a power-law like component, plus the lamppost-geometry relativistic reflection \citep[{\tt relxilllpCp},][]{2014Garc}, respectively. The galactic column density, $N_\mathrm{H}^\mathrm{Gal}$, is allowed to vary due to the discrepancy between $N_\mathrm{H}^\mathrm{Gal}=1.5\times10^{21}\,\mathrm{cm}^{-2}$ \citep{1990Dickey} and $N_\mathrm{H}^\mathrm{Gal}=1.06\times10^{21}\,\mathrm{cm}^{-2}$ \citep{2005Kalberla}. The inner radius of the reflection component is assumed at the innermost stable circular orbit (ISCO). Here we adopt a phenomenological model for the soft excess as its nature is still being debated \citep[e.g.][]{2006Crummy,2019Garc,2020Middei,2021XuSE}. The result of the time-averaged spectrum reveals the primary continuum with a slope of $\Gamma=2.15^{+0.01}_{-0.01}$, a soft excess characterized by a disk black-body with a temperature of $T_\mathrm{in}\sim0.13$\,keV, and a relativistic reflection component with a reflection fraction (a ratio of the intensity emitted towards the disk compared to that one escaping to infinity) of $f_\mathrm{Refl}=0.62^{+0.08}_{-0.04}$. The inclination angle, ionization state and the iron abundance (in units of Solar abundance) of the disk are required to be $i\sim39^{\circ}$, $\log(\xi/\mathrm{erg\,cm\,s^{-1})}=3.08^{+0.04}_{-0.04}$ and $A_\mathrm{Fe}=6.6^{+2.0}_{-1.7}$ respectively. The corona is measured at a height of $4.7^{+2.2}_{-2.3}\,R_\mathrm{Horizon}$ above the SMBH with a spin of $a_\star=0.21^{+0.35}_{-0.21}$ if we assume the corona as a compact point. All of these are similar to previous results \citep{2018Frederick,2020Waddell}. It is noticed that \citet{2019Jiang} adopted a high-density reflection model without {\tt diskbb} to fit the same data and obtained a high spin value of $a_\star>0.45$ with a similar abundance of $A_\mathrm{Fe}=5.9^{+0.6}_{-1.4}$ and a disk density of $\log(n_\mathrm{e}/\mathrm{cm}^{-3})\sim17.7$ (fixed at $\log(n_\mathrm{e}/\mathrm{cm}^{-3})=15$ in {\tt relxilllpCp}). We tried the same high-density model for our spectra and found that the fits are much poorer ($\Delta\chi^2\sim475$) than the current model, perhaps because they only consider the EPIC spectrum above 0.5\,keV, while we extend that to 0.4\,keV and the high-density reflection cannot explain that band well.

\begin{figure}
	\includegraphics[width=\columnwidth, trim={50 90 80 30}]{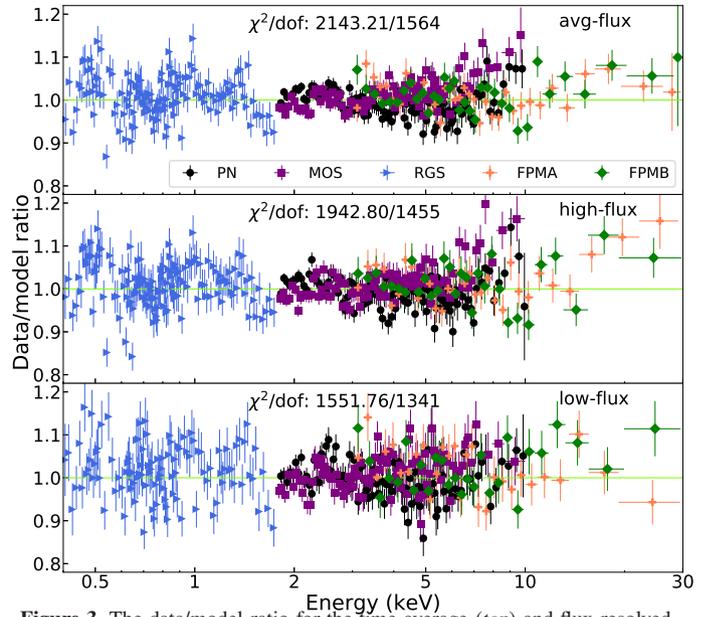}
    \caption{The data/model ratio for the time-average ({\it top}) and flux-resolved ({\it middle} and {\it bottom}) spectra with respect to the best-fit continuum model: {\tt tbabs*zashift*(diskbb+relxilllpCp)}. The $\chi^2$ statistics and the degree of freedom (dof) are marked as well. The residuals are binned for clarity.}
    \label{fig:ratio}
\end{figure}

The same best-fit model applies to the flux-resolved spectra, with several parameters linked to the time-average results, including the inclination angle, black hole spin, iron abundance, and the temperature of corona, which are not expected to vary over a few hours. The data/model ratio are depicted in Fig.\ref{fig:ratio} and the details of fits are shown in Tab.\ref{tab:fits}. As explained by \citet{2018Frederick}, the decreasing corona height and ionization parameter plus the increasing reflection fraction during the low-flux state could be attributed to the light-bending effect, where the corona may drop down closer to the accretion disk causing the dip in flux.

%%%%%%%%%%%%%%%%%%%%%%%%%%%%%%%%%%%%%%%%%%%%%%%%%%
%%%%%%%%%%%%%%%%%%%%%%%%%%%%%%%%%%%%%%%%%%%%%%%%%%

\subsection{Direct Gaussian Line Scan}\label{subsec:line-scan}
To search for any strong and narrow features upon the continuum, we perform a Gaussian line scan, which is a powerful tool to visualize the lines over the continuum in the \xmm\ energy band. We ignore \nustar\ data due to its lower spectral resolution. We fit a Gaussian line spanning the $0.4\mbox{--}10$\,keV range with a logarithmic grid of energy steps. The energy centroid and the line width are fixed at each step, and the normalization could vary to be positive or negative, in order to reproduce both emission and absorption lines. The grid of the line width $\sigma_v$ in km/s ranges from 100 to to and 10000\,km/s and the corresponding line width in keV is $\sigma_E=\frac{\sigma_v}{c}E$, where $c$ is the speed of light and $E$ is the energy. For the sake of the balance between the computational cost and the resolving power, we employ different numbers of points $N_v$ for each line width so that the product of $\sigma_v$ and $N_v$ equals to $10^{6}$. The parameters of the broadband model are left free during the scan. The $\Delta\chi^2$ improvement to the best-fit continuum model is recorded at each step. 

%%%%%%%%%%%%%%%%%%%%%%%%%%%%%%%%%%%%%%%%%%%%%%%%%%
\begin{figure}
	\includegraphics[width=\columnwidth, trim={50 90 30 30}]{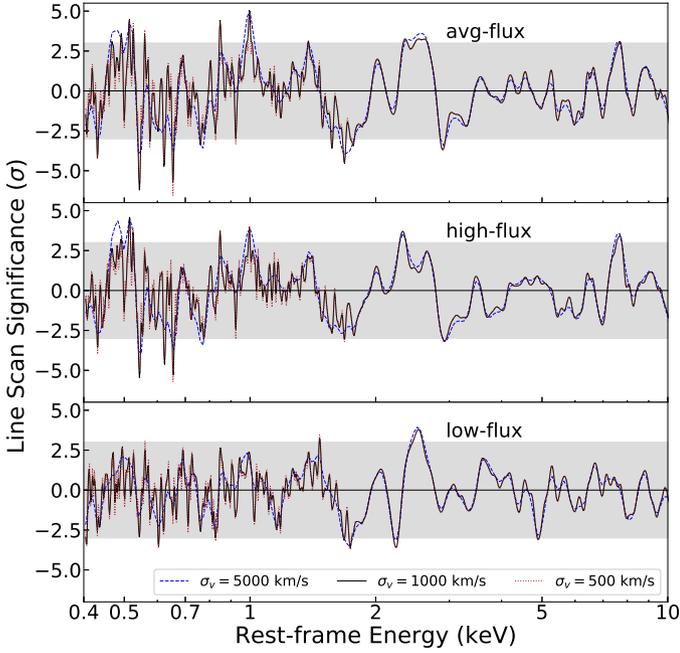}
    \caption{The results of the line search over the \xmm\ broadband time-average ({\it top}) and flux-resolved ({\it middle} and {\it bottom}) spectra in the AGN rest frame. The line width in velocity $\sigma_v$ ranges from 100 to 10000 km/s\yrxu{, while only 500, 1000, and 5000 km/s are shown for clarity}. The significance is expressed with the square root of $\Delta\chi^2$ times the sign of the Gaussian normalization. The {\it grey} region corresponds to a $3\sigma$ value of the single trial significance.}
    \label{fig:gaussian}
\end{figure}
%%%%%%%%%%%%%%%%%%%%%%%%%%%%%%%%%%%%%%%%%%%%%%%%%%
%%%%%%%%%%%%%%%%%%%%%%%%%%%%%%%%%%%%%%%%%%%%%%%%%%
\begin{figure}
	\includegraphics[width=\columnwidth, trim={30 50 50 50}]{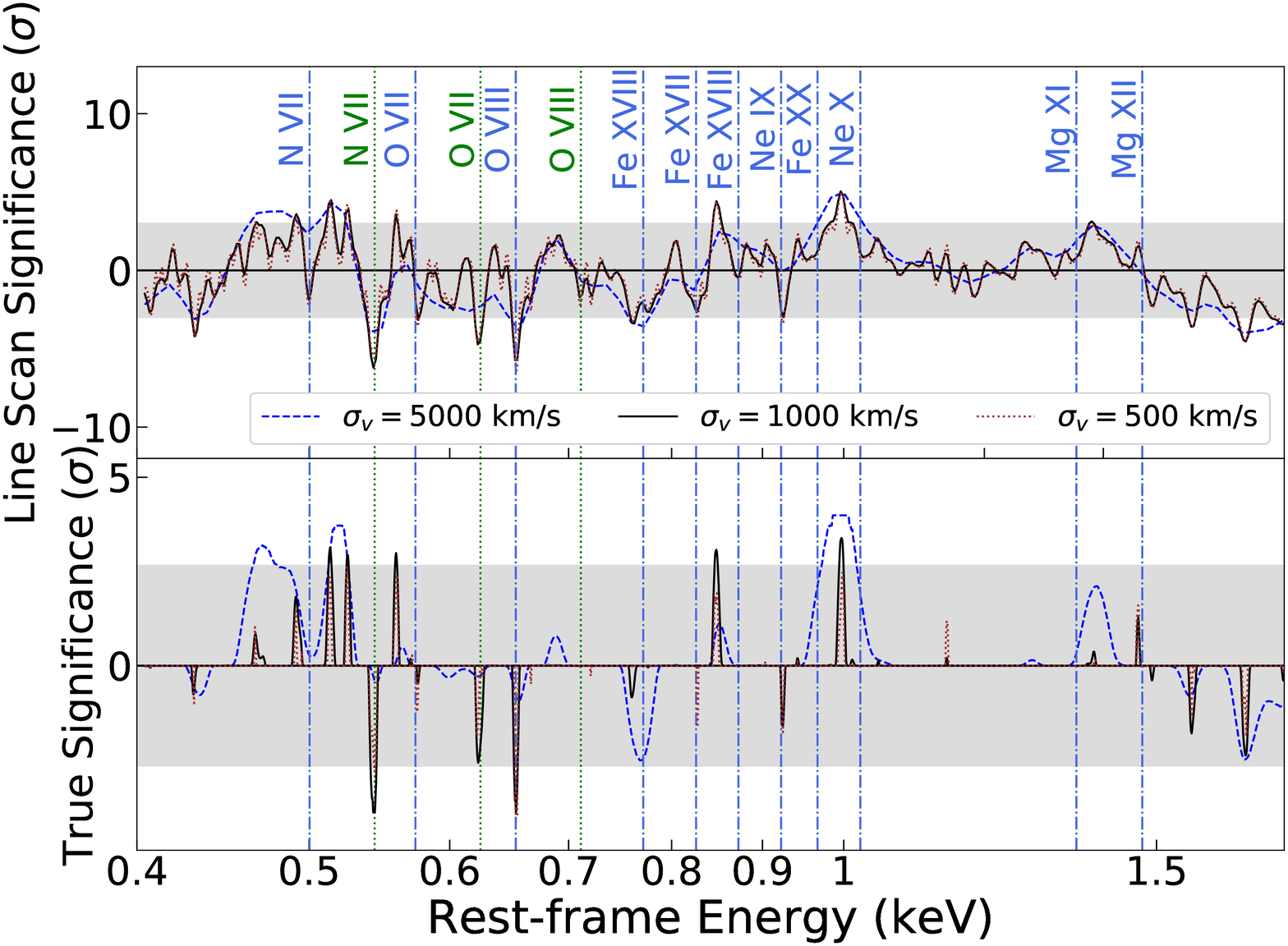}
    \caption{The single trial significance obtained from the Gaussian line scan with different line width over the RGS spectrum is displayed in the \yrxu{{\it top} panel}. The {\it grey} region marks the significance 3$\sigma$. The true significance of the residuals (\yrxu{{\it bottom}}) is obtained from the Monte Carlo simulations \yrxu{with the same choice of line widths}. The {\it grey} region represents 99\% significance level.}
    \label{fig:MC+Gauss}
\end{figure}
%%%%%%%%%%%%%%%%%%%%%%%%%%%%%%%%%%%%%%%%%%%%%%%%%%
The results of the scan over the time-averaged and flux-resolved spectra are expressed as the square root of $\Delta\chi^2$ times the sign of the normalization in Fig.\ref{fig:gaussian} \yrxu{(only results of 500, 1000, 5000 km/s line width are shown for clarity, which is the same reason for following plots)}. This quantity provides a rough estimate of the single trial detection significance of each Gaussian line (ignoring the look-elsewhere effect). The {\it grey} region marks the 3$\sigma$ significance level. The zoom-in result for the time-averaged spectrum over the RGS band is plotted in the \yrxu{top} panel of Fig.\ref{fig:MC+Gauss} for clarity. We mark the rest-frame positions of the main ionic transition lines within the \xmm\ band with the vertical \yrxu{{\it blue dashdot}} lines. Coincidentally, we identify several absorption lines at their rest-frame energies: N {\small VII}, O {\small VII}, O {\small VIII}, Fe {\small XVIII} and Ne {\small IX}. Other unidentified lines are also marked as the reference. As for the other more or less significant absorption lines, three of them could be identified as N {\small VII}, O {\small VII} and O {\small VIII} blueshifted by $\sim$23400\,km/s (corresponding to $z\sim-0.078$), presented by the vertical \yrxu{{\it green dotted}} lines. Compared with the absorption features, the emission lines seems to be blue- or red-shifted and are not easy to identify as emission from the same outflowing gas. In particular, there is a strong broad emission around 1\,keV, halfway between the rest-frame positions of Fe {\small XX} and Ne {\small X}. We note that the residuals appear stronger in the spectrum of the bright state likely as an artificial result of a twice higher number of counts. We however also find that the normalisation of the Gaussian lines appear slightly stronger in the high-flux spectrum, although the trend cannot be confirmed due to the limited statistics.

\subsection{Monte Carlo Simulation}\label{subsec:MC}
Although the Gaussian line scan is powerful in locating any possible spectral lines, many spurious features might appear significant. Owing to the look-elsewhere effect: searching a large parameter space could enhance the chances that it contains a strong feature originating from pure noise \citep{2008Vaughan}. To account for this effect, Monte Carlo (MC) simulations should be run to estimate the false positive rate of any detected feature. In principle, one must simulate a set of spectra with statistics comparable to the real data based on the best-fit broadband model only affected by Poisson noise. Then the Gaussian line scan is repeated on each simulated dataset to obtain the fraction of spectra with fake features stronger (i.e. with higher $\Delta \chi^2$) than the one detected in the real spectra. Therefore, if we want to confirm a $\sim4\sigma$ significance for the detected lines, the Gaussian line scan has to been performed over 10000 simulated datasets, which is computationally expensive.

For the sake of the computational costs, we adopt the new method recently developed by \citet{2021Kosec}, which employs the cross-correlation approach instead of fitting the Gaussian line spanning the given energy band. Since we are interested in the resolved soft X-ray features, the cross-correlation method is only applied to the time-average RGS spectrum. \yrxu{We employ a logarithmic grid of 2000 points among RGS energy coverage and line widths ranging from 0 to 5000 km/s. The cross-correlation is calculated in form of}
$$C=\sum\limits^N_{i=1}x_iy_i,$$
where $x$ and $y$ are respectively the arrays of the real residual spectrum and the Gaussian line model at a predefined width and centroid energy. \yrxu{The detailed procedures are described in Appendix.\ref{sec:CC-MC}.}

The true significance is reported in the bottom panel of Fig.\ref{fig:MC+Gauss}. The true significance confirms the existence the absorption lines including O {\small VIII}, blueshifted N {\small VII} and O {\small VII}. We also find five emission lines above the 99\% significance level, particularly the $\sim$1\,keV emission, which is also found in section~\ref{subsec:line-scan} and will be investigated in section~\ref{subsec:XSTAR-emission}. We caution that lines with a weak true significance are not necessarily fake as the true significance provides a very conservative approach by assuming the worst case of a single and independent line with any shift in the observed spectrum. The photoionization model is therefore required to model these lines simultaneously.

\subsection{Search for Outflows}\label{subsec:outflow}
In this section, we employ the photoionization code XSTAR \citep{2001Kallman}, to describe the absorption features in 1H 1934 spectrum and to investigate the outflow variability during the observation. XSTAR computes the physical conditions and synthetic spectra of the gas photoionized by a given radiation field. It calculates the intensity for a large number of lines, and enables fits to data through the constructed table of synthetic spectra. 

%%%%%%%%%%%%%%%%%%%%%%%%%%%%%%%%%%%%%%%%%%%%%%%%%%
\begin{figure}
	\includegraphics[width=\columnwidth, trim={50 50  50 0}]{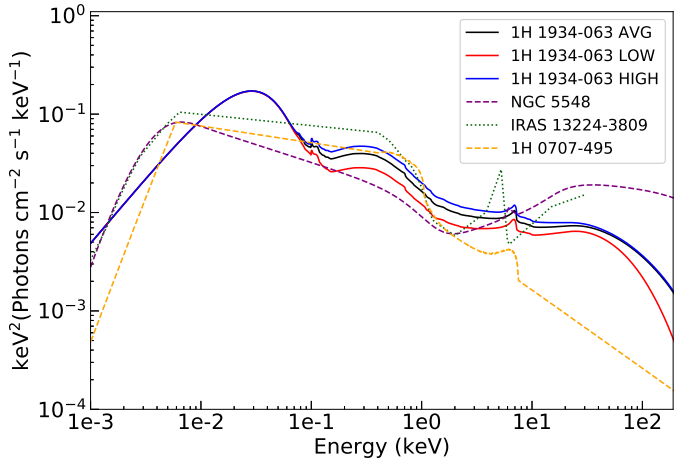}
    \caption{The time-averaged and flux-resolved SEDs of 1H 1934 compared with other Seyfert galaxies \yrxu{(NGC 5548, \citealt{2015Mehdipour}; IRAS 13224-3809, \citealt{2018Jiang}; 1H 0707-495, \citealt{2021Xu})}. The UV/optical fluxes are predicted by a color-corrected \citep[$E_\mathrm{B-V}=0.293$;][]{2006Riffel} additional {\tt diskbb} characterized by a temperature of $\sim12\,$eV.}
    \label{fig:SED}
\end{figure}
%%%%%%%%%%%%%%%%%%%%%%%%%%%%%%%%%%%%%%%%%%%%%%%%%%

The intrinsic spectral energy distribution (SED) of 1H 1934 input into XSTAR is computed from the UV/optical to hard X-ray energies by adding the OM data. The UV/optical fluxes are described by an additional {\tt diskbb} component characterized by a temperature of $\sim12\,$eV. The interstellar extinction \citep[$E_\mathrm{B-V}=0.293$;][]{2006Riffel} has been taken into account. The time-averaged and flux-resolved SEDs are shown in Fig.\ref{fig:SED} as compared with other Seyfert galaxy SEDs, where 1H 1934 SEDs look similar with that of IRAS 13224-3809. Accordingly, we estimate the accretion rate by measuring the bolometric luminosity from the time-averaged SED ($10^{-5}\mbox{--}10^{3}$\,keV), which is $L_\mathrm{bol}\sim1.725\times10^{44}\,\mathrm{erg}$. The accretion rate is thus high since we obtain $\dot{m}=L_\mathrm{bol}/L_\mathrm{Edd}\sim0.40^{+0.91}_{-0.27}$ if we adopt a SMBH mass of $3\times10^6\,M_\odot$ with a typical $\sim$0.5\,dex uncertainty \citep{2000Rodriguez-Ardila,2008Malizia}, where $L_\mathrm{Edd}=4\pi GM_\mathrm{BH}m_\mathrm{p}c/\sigma_T$ is the Eddington luminosity. 

For the modelling of the outflows in flux-resolved spectra, we still employ the XSTAR model generated by the time-averaged SED because the SED does not significantly vary during the observation (see Fig.\ref{fig:SED}). The input SED ranges from 1\,eV to 30\,keV to avoid the possible effects from $>30$\,keV band, which cannot be constrained with the available data. We adopt a grid that has a turbulence velocity of 100\,km/s (due to narrow lines detected in Fig.\ref{fig:gaussian}) and covers a wide range of ionization states ($\log(\xi/\mathrm{erg\,cm\,s^{-1})}=0\mbox{--}6$) and column densities ($N_\mathrm{H}\,\mathrm{(cm^{-2})}=10^{19}\mbox{--}10^{24}$) to compute the XSTAR table model. The code returns one emission and one absorption spectrum table, and here we use the absorption component ({\tt xstar\_abs}) for modelling. In next section~\ref{subsec:XSTAR-emission} we will use the emission component ({\tt xstar\_em}) to the model the emission features.

\subsubsection{Best-fit Outflow Model}\label{subsec:XSTAR-fits}
{\tt xstar\_abs} is a multiplicative model, characterized by the column density $N_\mathrm{H}$, the ionization parameter $\xi$ and the line-of-sight (LOS) redshift $z_\mathrm{LOS}$. To locate the global best-fit solution, we build a systematic scan over a large grid of the parameter space before directly fitting to the spectra \citep[e.g.][]{2020Kosec,2021Xu}. We create a multi-dimension grid of the $\log\xi$ ($0\mbox{--}6$ with $\Delta\log\xi=0.12$) and $z_\mathrm{LOS}$ (from $-0.25$ to $0.02$ with $\Delta z_\mathrm{LOS}=0.001$), and allow $N_\mathrm{H}$ to be free. The search is performed by adopting the broadband continuum obtained in section~\ref{subsec:continuum} and the parameters of the continuum model are left free. At each grid point, the statistical improvement $\Delta\chi^2$ is recorded, which presents the significance of the wind absorption at this point over the continuum model. One advantage of this method is to reveal the possible existence of the multiple phases in the outflow as it searches in a large parameter space.

\begin{figure}
	\includegraphics[width=\columnwidth, trim={30 10 50 0}]{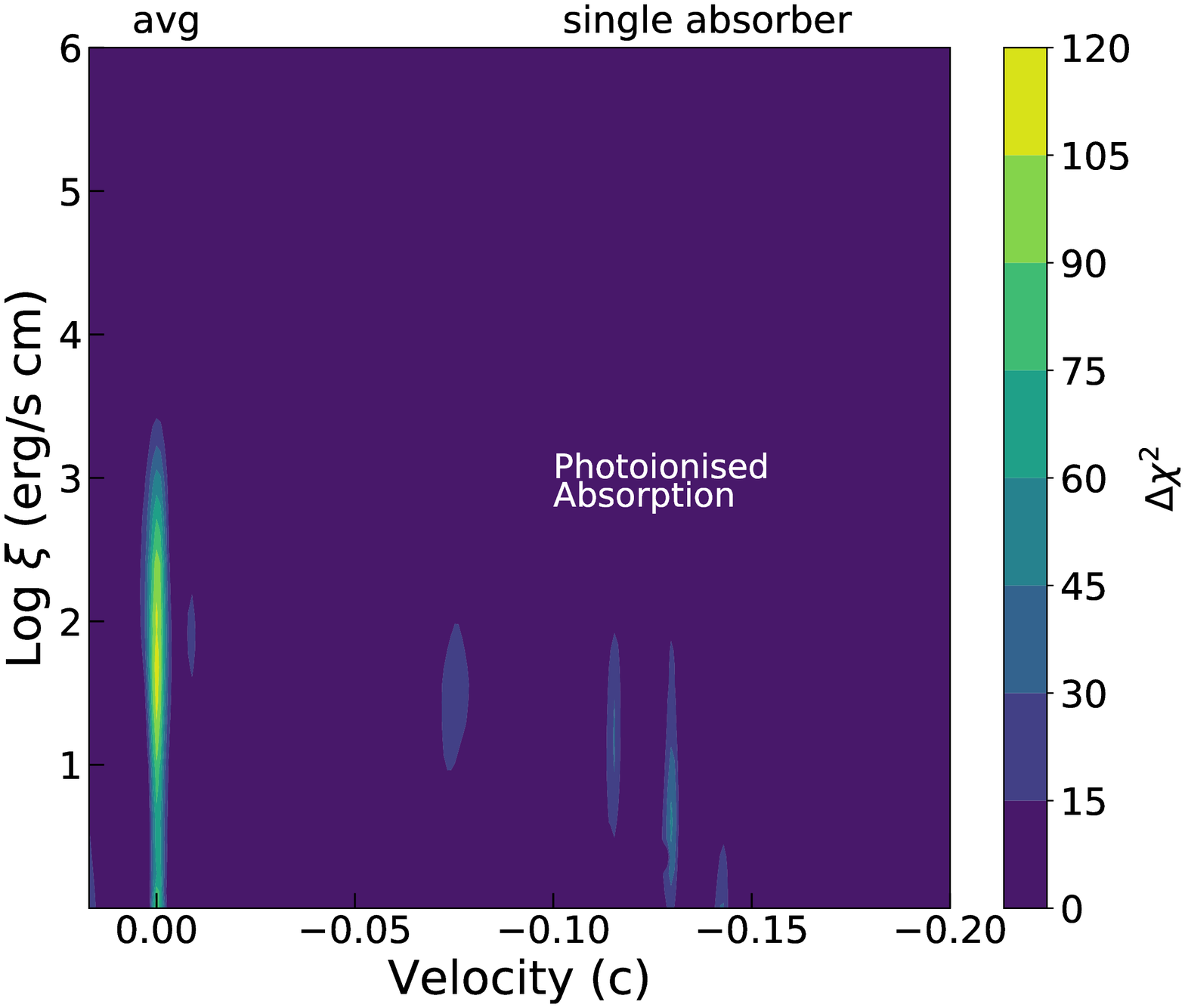}
	\includegraphics[width=\columnwidth, trim={30 20 50 0}]{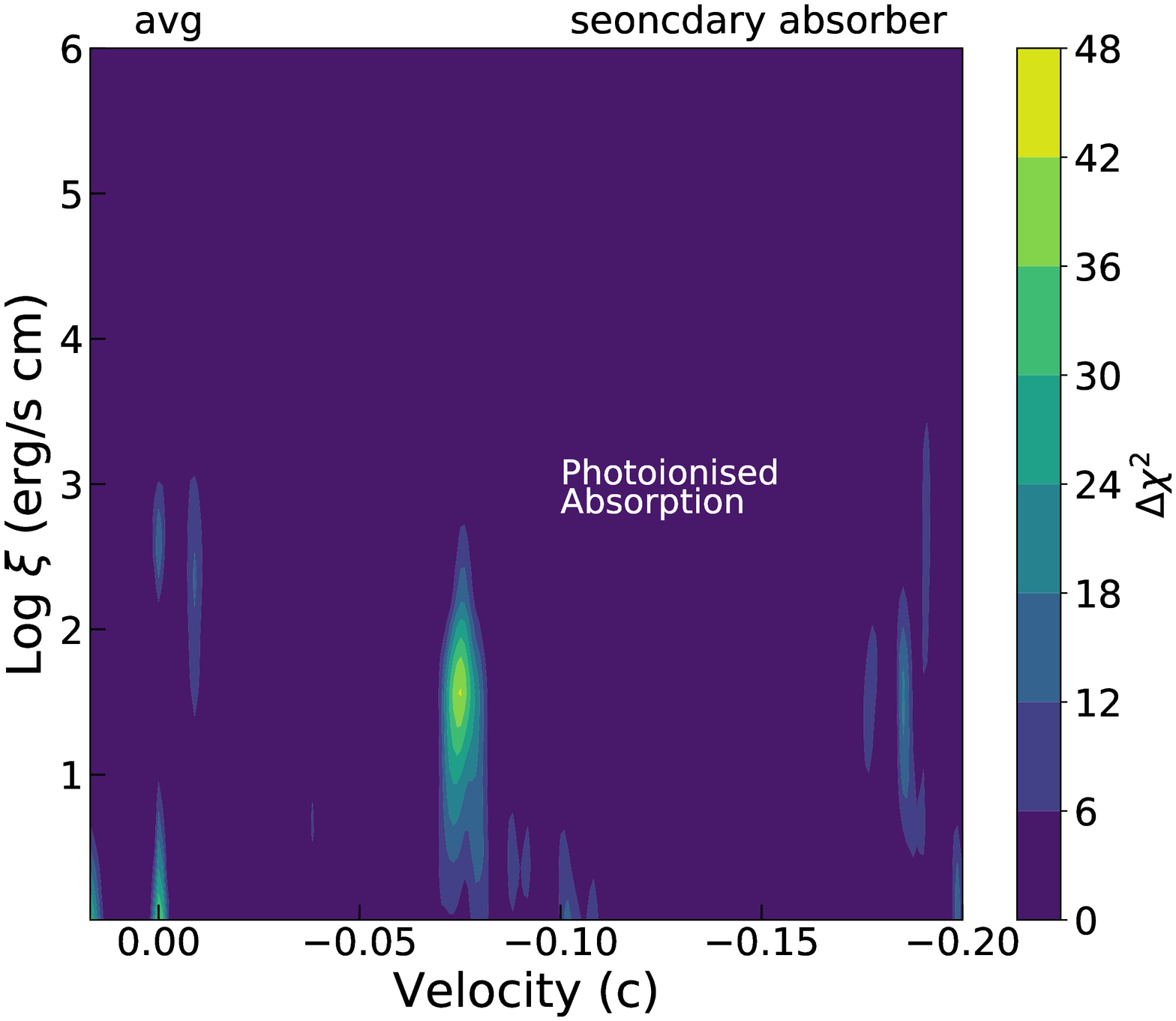}
    \caption{Photoionization absorption model search for the time-averaged spectrum of 1H 1934 over the broadband continuum model ({\it Upper}) and with the addition of a warm absorber ({\it Lower}) in the AGN rest frame. The velocity on the X-axis is relativistically corrected. The color illustrates the statistical improvement after adding a absorbing model.}
    \label{fig:xstar-scan}
\end{figure}

The scan result for the time-averaged spectrum is shown in the upper panel of Fig.\ref{fig:xstar-scan}. The line-of-sight redshift is relativistically corrected through equation: $v/c=\sqrt{(1+z)/(1-z)}-1$. The search achieves a very strong detection ($\Delta\chi^2\sim120$) of a mildly ionized ($\log(\xi/\mathrm{erg\,cm\,s^{-1})}\sim1.7$) absorber at the rest frame of AGN ($v\sim0$). The spectral fit reveals the column density of $N_\mathrm{H}=2.0^{+0.5}_{-0.4}\times10^{20}\,\mathrm{cm}^{-2}$. The systematic velocity and ionization parameter suggest that this solution corresponds to a common warm absorber as those found in other AGN \citep[e.g.][]{2011Detmers,2014Kaastra}. It is noted that there are several secondary peaks ($v<-0.05\,c$) in the plot, indicating the presence of a further faster ionized outflow component in the spectrum. However, the secondary peaks might not be independent of the primary warm absorber as they might explain the same residuals with different models.

%%%%%%%%%%%%%%%%%%%%%%%%%%%%%%%%%%%%%%%%%%%%%%%%%%
%%%%%%%%%%%%%%%%%%%%%%%%%%%%%%%%%%%%%%%%%%%%%%%%%%

We therefore perform a further {\tt xstar\_abs} scan over a new baseline model, which includes the best-fitting rest-frame warm absorber component and the broadband continuum. The searched parameter space remains the same and all the baseline model parameters are allowed to vary freely. The result is shown in the lower panel of Fig.\ref{fig:xstar-scan}. There is a less strong but still significant detection ($\Delta\chi^2=48$) of a secondary component, which is an ultra-fast (relativistically corrected velocity, $v\sim-0.075\,c$) outflow with a similar ionization parameter ($\log(\xi/\mathrm{erg\,cm\,s^{-1})}\sim1.6$). With such high $\Delta\chi^2$ with the physical models, Monte Carlo simulations are redundant as the detection will be highly significant. The fitting results of two {\tt xstar\_abs} components over the time-averaged spectrum are presented in the third column of Tab.\ref{tab:fits}. The column density of the UFO ($N_\mathrm{H}\sim6.6\times10^{19}\,\mathrm{cm}^{-2}$) is significantly lower than that of the warm absorber, resulting in weaker spectral features. The $\Delta\chi^2$ of including each absorption component is consistent with the original value found in the automated wind search, suggesting that the UFO and warm absorber do not fit the same spectral features. The model comparison among the continuum, plus one, and plus two absorber components are shown in the top panel of Fig.\ref{fig:flux-spectrum}. As expected, the spectral features of the warm absorber are mainly located at the rest frame of N {\small VII}, O {\small VII} and O {\small VIII} 1s-2p lines, while the main absorption lines of the UFO are the blueshifted N {\small VII} and O {\small VII} and a weaker O {\small VIII}, which are also marked in Fig.\ref{fig:MC+Gauss}. Such differences illustrate that the best-fit value of the UFO ionization parameter might be lower than that of the warm absorber, although it cannot be confirmed within the uncertainties.

\subsubsection{Outflow variability}\label{subsec:variability}

%%%%%%%%%%%%%%%%%%%%%%%%%%%%%%%%%%%%%%%%%%%%%%%%%%
\begin{table*}
\centering
\caption{Best-fit parameters of the 1H 1934-063 time-average and flux-resolved spectra modelling.
}
\begin{tabular}{lccccc}
\hline
\hline
Description & Parameter & avg & low-flux & high-flux\\
\hline
{\tt tbabs} &    $N^\mathrm{Gal}_\mathrm{H}$ ($10^{21}$ cm$^{-2}$)      & $2.1^{+0.1}_{-0.1}$ & $2.1^{t}$    &  $2.1^{t}$  \\
\hline
{\tt diskbb}      & $T_\mathrm{in}$ (keV)  &  $0.130^{+0.005}_{-0.005}$    & $0.129^{+0.004}_{-0.004}$ & $0.133^{+0.003}_{-0.003}$ \\
                &  $N_\mathrm{BB}$ ($10^{4}$)     &     $1.2^{+0.4}_{-0.3}$     & $0.9^{+0.2}_{-0.2}$ & $1.3^{+0.2}_{-0.2}$   \\
\hline
{\tt relxilllpCp}   & $h$ ($R_\mathrm{Horizon}$)  & $4.7^{+2.2}_{-2.3}$ & $<2.9$ & $6.4^{+3.2}_{-2.4}$ \\
                    & $a_\star$ ($cJ/GM^2$) & $0.21^{+0.35}_{-0.21}$  & $0.21^{t}$ & $0.21^{t}$ \\
                    & $i$ (deg) & $39.0^{+1.5}_{-1.2}$ & $39.0^{t}$ & $39.0^{t}$ \\	
                    & $\Gamma$   &  $2.15^{+0.01}_{-0.01}$  &  $2.09^{+0.01}_{-0.01}$  & $2.18^{+0.01}_{-0.01}$  \\
                    & $\log(\xi/\mathrm{erg\,cm\,s^{-1})}$ & $3.08^{+0.04}_{-0.04}$ & $3.03^{+0.05}_{-0.06}$ & $3.12^{+0.06}_{-0.05}$ \\
                    &  $A_{\mathrm{Fe}}$ & $6.6^{+2.0}_{-1.7}$ & $6.6^{t}$ & $6.6^{t}$ \\
                    & $kT_\mathrm{e}$ (keV) & $400^{+0}_{-317}$ & $400^{t}$ & $400^{t}$ \\
                    & $f_\mathrm{Refl}$ &  $0.62^{+0.08}_{-0.04}$ & $0.84^{+0.08}_{-0.10}$  & $0.53^{+0.07}_{-0.07}$  \\
                    &  $N_\mathrm{refl}$ ($10^{-3}$)     & $0.19^{+0.14}_{-0.04}$ & $2.4^{+8.6}_{-2.1}$  & $0.19^{+0.05}_{-0.02}$  \\
\hline
  broadband        &  $\chi^{2}$/d.o.f.(degree of freedom) & 2143.21/1564 & 1551.76/1341 & 1942.81/1455  \\
\hline
{\tt xstar\_abs1}     & $N_\mathrm{H}$ ($10^{20}$\,cm$^{-2}$)   &  $2.0^{+0.5}_{-0.4}$  & $5.9^{+2.5}_{-2.6}$ & $2.1^{+0.6}_{-0.5}$ \\
(warm absorber)        &    $\log(\xi/\mathrm{erg\,cm\,s^{-1})}$   &  $1.68^{+0.12}_{-0.20}$ & $2.42^{+0.09}_{-0.13}$ & $1.71^{+0.06}_{-0.18}$ \\
        &    $z_\mathrm{LOS}$ ($10^{-4}$) &  $-0.78^{+3.6}_{-3.6}$ & $-2.9^{+6.5}_{-7.7}$ & $-0.1^{+3.7}_{-4.5}$  \\
\hline
  broadband+abs1      &  $\chi^{2}$/d.o.f. & 2028.50/1561 & 1517.82/1338 & 1857.47/1452  \\
\hline
{\tt xstar\_abs2} & $N_\mathrm{H}$ ($10^{19}$\,cm$^{-2}$)   &  $6.6^{+4.1}_{-2.1}$  & $7.9^{+6.4}_{-4.3}$ & $5.1^{+3.2}_{-2.3}$ \\
(UFO)        &    $\log(\xi/\mathrm{erg\,cm\,s^{-1})}$    &  $1.6^{+0.1}_{-0.1}$ & $1.8^{+0.4}_{-0.4}$ & $1.5^{+0.2}_{-0.3}$ \\
        &    $z_\mathrm{LOS}$ ($10^{-2}$) &  $-7.75^{+0.09}_{-0.13}$ & $-7.75^{+0.26}_{-0.19}$ & $-7.83^{+0.17}_{-0.38}$ \\
\hline
 broadband+abs1+abs2         &   $\chi^{2}$/d.o.f. & 1985.14/1558 & 1501.62/1335 & 1833.95/1449  \\
 \hline
\end{tabular}
\label{tab:fits}
\begin{flushleft}
{$^{t}$ The parameter is tied.}
\end{flushleft}
\end{table*}
% %%%%%%%%%%%%%%%%%%%%%%%%%%%%%%%%%%%%%%%%%%%%%%%%%%%%%%%%%%%%%%%%%%%%%%

\begin{figure}
	\includegraphics[width=\columnwidth, trim={50 50  50 0}]{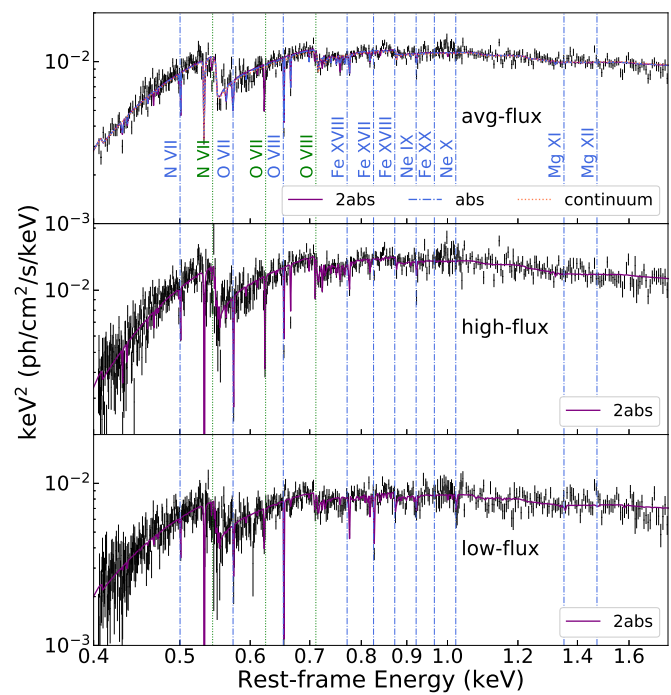}
    \caption{The best-fit of the continuum plus two absorber model ({\it purple \yrxu{solid}}) with respect to the folded time-averaged ({\it Top}) and flux-resolved ({\it Middle} and {\it Bottom}) X-ray spectrum of 1H 1934. Only the RGS spectrum is shown for clarity. The top panel also contains the fits with the continuum model ({\it coral \yrxu{dotted}}) and the continuum in addition to one absorption component ({\it blue \yrxu{dashdot}}). }
    \label{fig:flux-spectrum}
\end{figure}

As for the flux-resolved spectra, we directly fit them with the best-fit model (continuum plus two absorbers), and the parameters and spectra are shown in Tab.\ref{tab:fits} and Fig.\ref{fig:flux-spectrum} respectively. The statistic improvement of the high-flux spectrum is larger than that of the low-flux one, which is compatible with what we found in the line scan result (see section~\ref{subsec:line-scan}). The stronger residuals in the high-flux state might be also justified by the lower ionization state of the warm absorber and UFO, implying that fewer ions are over-ionized in the RGS band. Although the UFO could be regarded as stable during the observation within the uncertainties, the warm absorber slightly varies from the low- to high- flux state in the column density ($N_\mathrm{H} (10^{20}\,\mathrm{cm}^{-2})$ from $5.9^{+2.5}_{-2.6}$ to $2.1^{+0.6}_{-0.5}$) and the ionization parameter ($\log\xi$ from $2.42^{+0.09}_{-0.13}$ to $1.71^{+0.06}_{-0.18}$) at 90\% confidence level. It would be interesting if this is real and this variability comes from the response to the change in source luminosity (or to the interaction of the UFO with the surrounding medium - should the warm absorber represent the UFO shock front), as the warm absorber is usually expected to be stable far from the central SMBH and the response timescale should be much larger than dozens of hours. 

%%%%%%%%%%%%%%%%%%%%%%%%%%%%%%%%%%%%%%%%%%%%%%%%%%
%%%%%%%%%%%%%%%%%%%%%%%%%%%%%%%%%%%%%%%%%%%%%%%%%%

%%%%%%%%%%%%%%%%%%%%%%%%%%%%%%%%%%%%%%%%%%%%%%%%%%
\begin{figure}
	\includegraphics[width=\columnwidth, trim={50 10  50 0}]{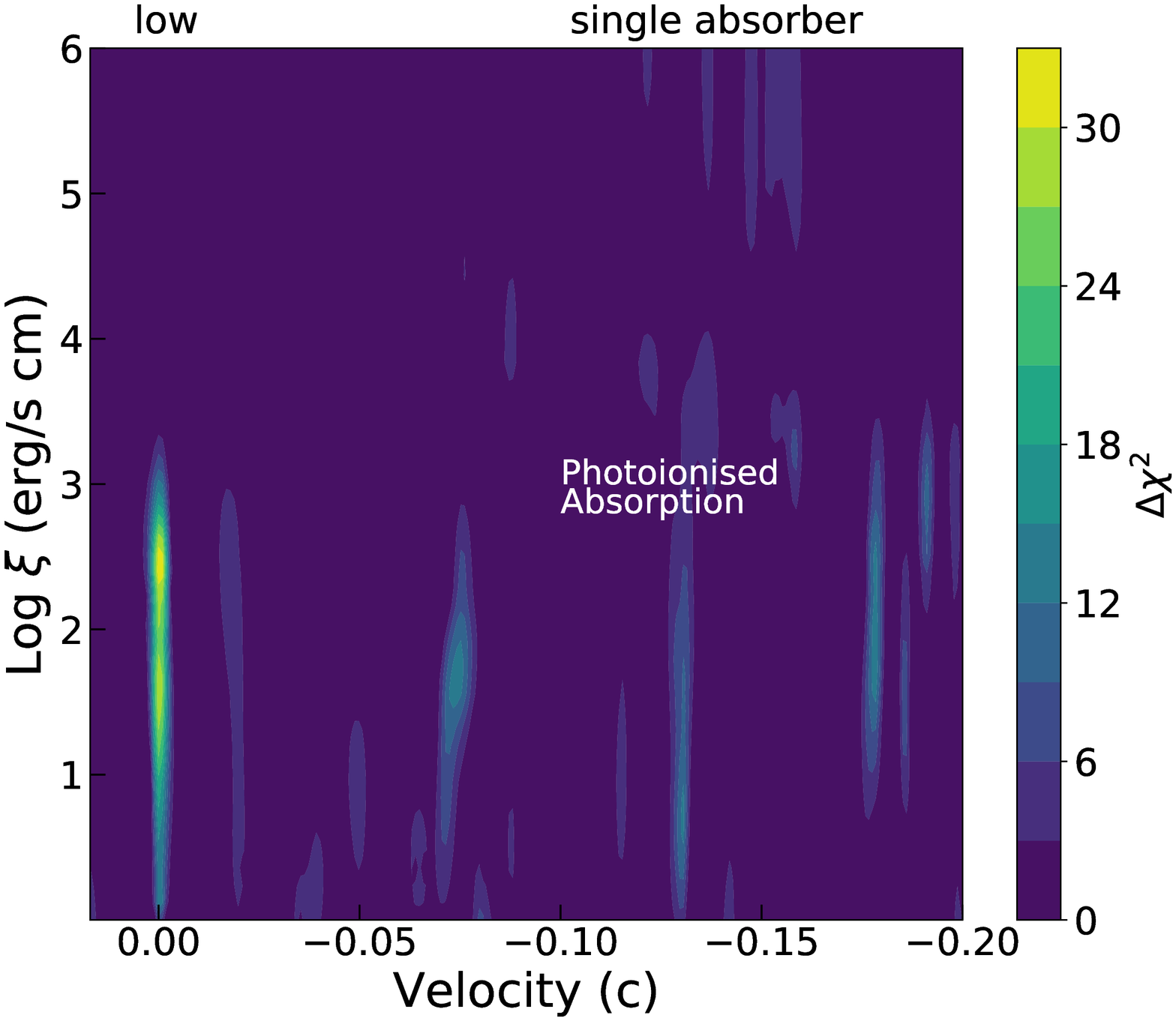}
	\includegraphics[width=\columnwidth, trim={50 20  50 0}]{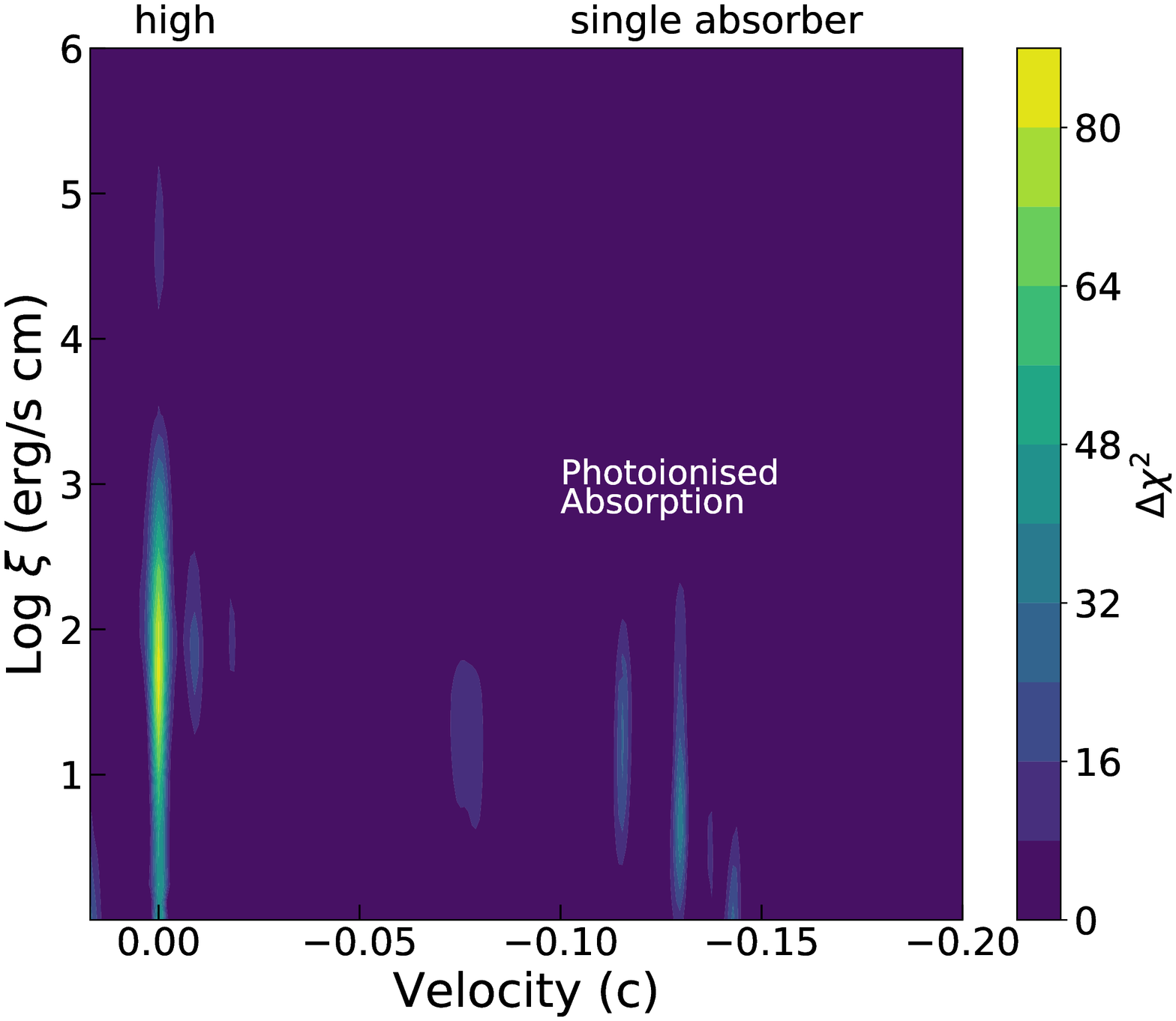}
    \caption{Photoionization absorption model search for the low- ({\it top}) and high-flux ({\it bottom}) spectra of 1H 1934 over the continuum model, similar with Fig.\ref{fig:xstar-scan}, with a single photoionized absorber.}
    \label{fig:variable-scan}
\end{figure}
%%%%%%%%%%%%%%%%%%%%%%%%%%%%%%%%%%%%%%%%%%%%%%%%%%

To investigate the validity of this discovery, we perform the same wind search on the flux-resolved spectra over the broadband continuum, of which results are shown in Fig.\ref{fig:variable-scan}. The ionization parameter of the primary solution (i.e. the warm absorber) in both spectra ranges from 0 to 3, suggesting that the observed variation might be artificial if we slightly loosen the confidence level of the parameters. However, the possibility of a variable warm absorber cannot be excluded without more observations. 
%Furthermore, the UFO detection obviously appears in the low-flux plot, but it is hard to ascribe its appearance to the non-fully-ionized UFO at the low-flux state \citep[e.g.][]{2018Pinto,2021Xu}, or to the less significance ($\Delta\chi^2\sim34$) of the warm absorber so that the UFO detection is comparable and could be observed, or their combination. 
The statistical improvements of the UFO detection in low- and high-flux spectra are comparable ($\Delta\chi^2\sim16$ and $\sim24$ respectively), rather than the dramatic change in those of the warm absorber ($\Delta\chi^2\sim34$ and $\sim85$ separately). The main change in the outflow during the exposure therefore seems to originate from the warm absorber. More observations are needed to confirm this.

\subsection{Emission Line Modelling}\label{subsec:XSTAR-emission}
In this section, we explore the nature of the unknown emission, especially for the line around 1\,keV, in the time-averaged spectrum of 1H 1934 for enough counts. Compared to the absorption, the emission is more complex and highly depends on the geometry, as it collects all photons that are not directed along the LOS towards the source. None of the emission lines detected in Fig.\ref{fig:MC+Gauss} are at the rest-frame of any known strong photoionization emission lines, implying that either they are Doppler-shifted photoionization lines or they are produced by other processes. The baseline model used in this section is the broadband continuum plus two absorbers adopted in section\,\ref{subsec:outflow} and Table~\ref{tab:fits}.
 
\subsubsection{Photoionization Emission}\label{subsubsec:photon-emission}
\begin{figure}
	\includegraphics[width=\columnwidth, trim={30 30  50 0}]{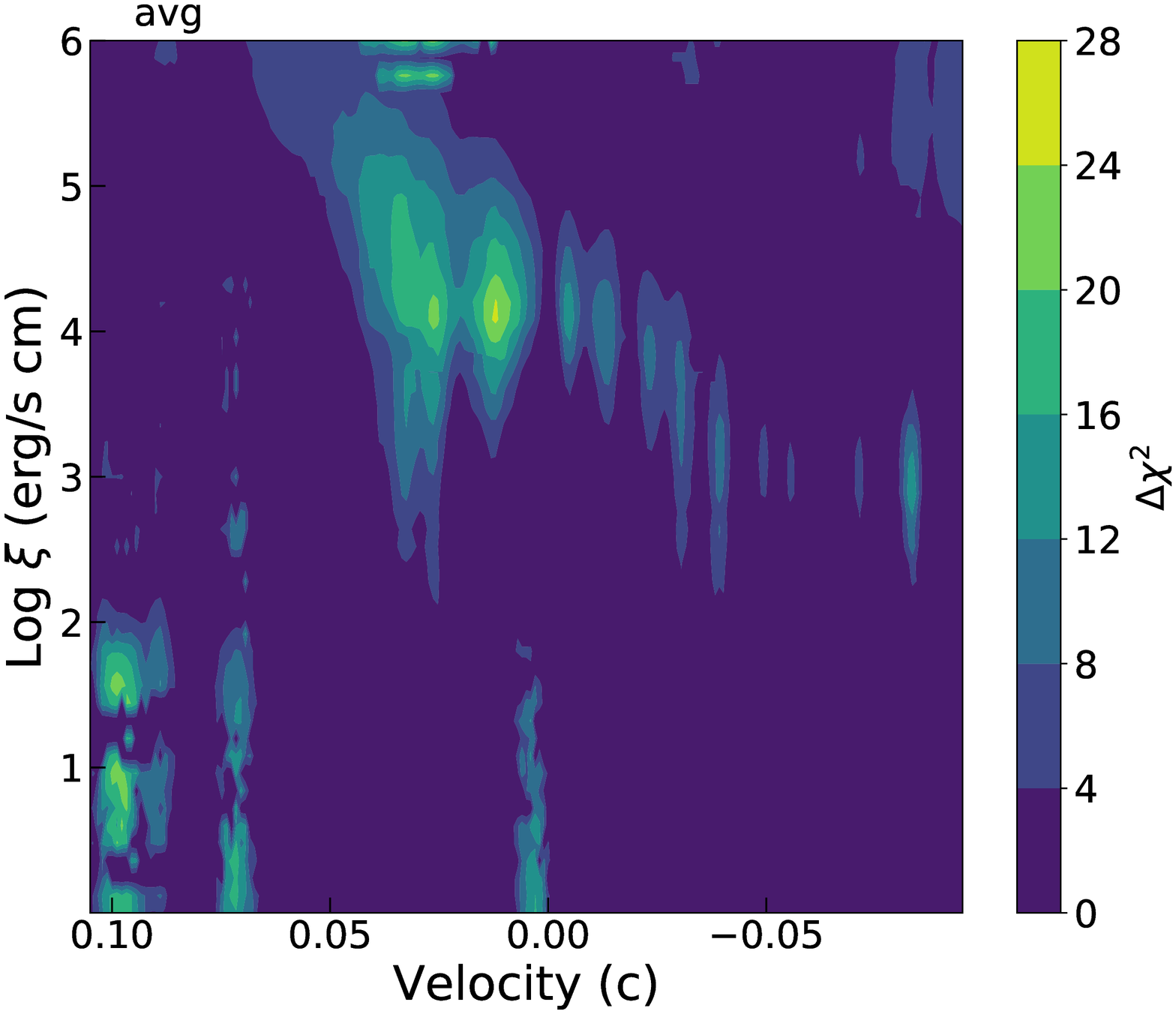}
    \caption{Photoionization emission model search for the time-averaged spectrum for the broadband model plus two absorbers, similar to Fig.\ref{fig:xstar-scan}. }
    \label{fig:EM-scan}
\end{figure}

It is natural to expect photoionized emission in the spectrum after finding two photoionization absorption components. We hence launch a physical emission model scan on the spectrum fitted with the new baseline model, where we adopt the XSTAR emission spectrum table ({\tt xstar\_em}) generated in section~\ref{subsec:outflow}. The range of the searched grid of the ionization parameter is the same as that of the absorption component, while the redshift spans between $-0.1$ and $0.1$ as we are not sure whether the emission lines are blue- or red-shifted and do not expect a high velocity of emitting gas. The results is shown in Fig.\ref{fig:EM-scan}, presenting a series of peaks at different shifts with weak detection statistics.

%%%%%%%%%%%%%%%%%%%%%%%%%%%%%%%%%%%%%%%%%%%%%%%%%%
%%%%%%%%%%%%%%%%%%%%%%%%%%%%%%%%%%%%%%%%%%%%%%%%%%

The best-fit solution ($\Delta\chi^2\sim28$) of {\tt xstar\_em} requires a column density of $N_\mathrm{H}<3.3\times10^{23}\,\mathrm{cm}^{-2}$, ionization parameter $\log(\xi/\mathrm{erg\,cm\,s^{-1})}=4.1^{+0.2}_{-0.1}$, and redshift $z_\mathrm{LOS}=1.5^{+5.2}_{-11.4}\times10^{-4}$. The velocity is consistent with the rest frame, and the ionization parameter is over two orders of magnitude higher than that of the absorption component, indicating that the light elements are fully ionized and the emission may originate from a different plasma than the absorbing gas. The best-fit spectrum is shown in the top panel of Fig.\ref{fig:em_comparison}, where the contribution of {\tt xstar\_em} is negligible in RGS band due to the high ionization state. We have checked that {\tt xstar\_em} mainly explains the high-ionization features such as Fe {\small XXV/XXVI}. Even if we add another photoionization emission component, the secondary emitter only explains some putative redshifted ($z_\mathrm{LOS}\sim0.0024$) O {\small VII} lines at a low ionization state ($\log(\xi/\mathrm{erg\,cm\,s^{-1})}<1.6$) with an even weaker improvement ($\Delta\chi^2\sim13$). Regardless of the detection significance and plausibility of the highly photoionized emission, it seems that photoionization cannot well model the broad line around 1\,keV, even if we tried using a convolution model ({\tt gsmooth}) to broaden the line width of {\tt xstar\_em}.

\subsubsection{Collision Ionization Emission}\label{subsubsec:collision-emission}
Collision ionization provides an alternative origin for the emission lines, since it could predict emission around 1 keV (Fe L-shell transitions) stronger than photoionisation \citep{2002Kinkhabwala}. We adopt the well-known {\tt bvapec} model in XSPEC with Solar abundance. This reproduces the emission spectrum of a plasma in collisional ionization equilibrium (CIE), which is characterized by its temperature and turbulent velocity. The best-fit model requires a temperature of $kT_\mathrm{e}^\mathrm{CIE}=1.18^{+0.11}_{-0.09}\,$keV, a velocity broadening of $\sigma_v^\mathrm{CIE}<7500$\,km/s, and a redshift of $z_\mathrm{LOS}^\mathrm{CIE}=0.011^{+0.001}_{-0.001}$. The corresponding spectrum is presented in the middle panel of Fig.\ref{fig:em_comparison}, revealing a series of emission lines around 1\,keV. The strongest and over-predicted line is Fe {\small XXI} with a rest-frame wavelength of 12.286\,{\AA} (i.e. 1.009 keV). A fit with a free Fe abundance does not provide better statistics. The remarkable fit improvement ($\Delta\chi^2\sim42$) suggests that the collision ionization seems to be a promising explanation for the 1\,keV emission but overpredicts emission of 1 keV.

\subsubsection{Emission of a secondary reflector}\label{subsubsec:reflection-emission}
Fluorescence from the disk-reprocessing of the inner coronal photons is another potential origin, as the time-averaged spectrum is dominated by the reflection ($f_\mathrm{Refl}\sim0.6$). The reflection lines in a rotational disk are anticipated to have asymmetric double peaks where the blue wing is stronger due to the relativistic beaming. 

Initially, we utilized a phenomenological model, {\tt diskline}, to describe line emission from a relativistic accretion disk. We adopted two disk line components to model the line around 1\,keV and the possible N {\small VII} emission at 0.5\,keV, because of the indicative wings around these two energies (see Fig.\ref{fig:MC+Gauss}). The inclination angle is linked to the primary reflection component angle. The outer radius is fixed at 1000\,$R_\mathrm{g}$ and the inner disk radius of {\tt diskline} are tied together as we do not expect their positions to be significantly different. The fit returns a $\Delta\chi^2\sim80$ improvement with a inner radius of $56^{+10}_{-11}\,R_\mathrm{g}$. The emissivity index is around 3, consistent with the solution in Newtonian spacetime. The best-fit centroid of the higher energy line is at $0.996^{+0.027}_{-0.046}\,$keV, which agrees with the Gaussian line fit. 

%%%%%%%%%%%%%%%%%%%%%%%%%%%%%%%%%%%%%%%%%%%%%%%%%%
\begin{figure}
	\includegraphics[width=\columnwidth, trim={30 50 50 0}]{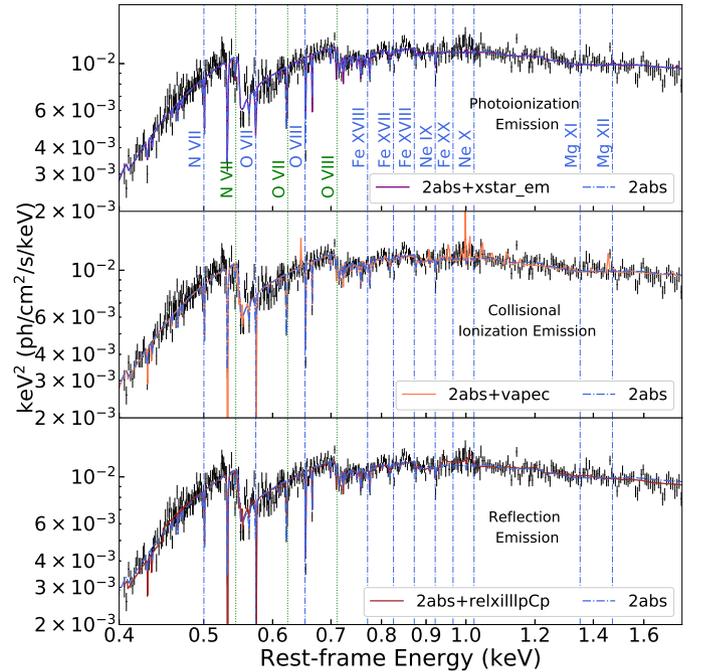}
    \caption{Fits with different models for emission, including photo- ({\it Top}), collision-ionization ({\it Middle}), and reflection ({\it Bottom}) for the time-averaged RGS spectrum. The baseline model is the broadband continuum plus two absorption components (see Tab.\ref{tab:fits}).}
    \label{fig:em_comparison}
\end{figure}
%%%%%%%%%%%%%%%%%%%%%%%%%%%%%%%%%%%%%%%%%%%%%%%%%%

Then we replace the phenomenological disk line models with a physical model, another {\tt relxilllpCp} component. The parameters are linked to those of the first {\tt relxilllpCp} component, except for the free inner disk radius, ionization parameter and normalization. The redshift is also allowed to vary, because the strongest reflection line in the soft band is O {\small VIII} and we thus tested whether such a strong 1\,keV emission could be the blueshifted O {\small VIII}, which is plausible for the reflection off the inner region of a rapidly rotational disk or the base of the wind. The best fit reveals a moderately ionized reflector with a strong blueshift of $z^\mathrm{Rel}_\mathrm{LOS}\sim-0.31$, resulting in a small $\Delta\chi^2\sim32$ improvement. A reflection component that is blueshifted by $\sim$0.3 would predict a strong broad iron line at $\sim8$\,keV. Such a feature is not observed in the spectrum, and thus the normalization of this component is low, and thus under-predicts the 1 keV line, as well.

However, the fit highly improves if we free the spectral slope in the second reflection model. A much softer ($\Gamma>3.2$) continuum irradiating the secondary reflector results in a remarkable statistical improvement of $\Delta\chi^2\sim75$, requiring a blueshifted ($z_\mathrm{LOS}^\mathrm{Rel}=-0.332^{+0.002}_{-0.002}$) ionized ($\log(\xi/\mathrm{erg\,cm\,s^{-1})}=2.50^{+0.25}_{-0.11}$) reflector with an inner radius of $R_\mathrm{in}=186^{+66}_{-106}\,R_\mathrm{g}$. The best-fit RGS spectrum is illustrated in the bottom panel of Fig.\ref{fig:em_comparison}. The main contribution of the secondary reflector to the modelling is the flux around 1\,keV with a double peak profile, which is produced by distant inner radius. The requirement for such a soft continuum implies that the emission lines are from a gas irradiated by both the hot corona and the soft excess, no matter whether the soft excess originates from a warm corona or the relativistic reflection. We actually attempted to replace the {\tt diskbb} with a Comptonization model {\tt nthComp} and adopted this as the radiation field of the second reflector, to investigate the origin of the X-ray radiation field. We did not find any significant improvement with respect to the disk model and thus preferred to keep the {\tt diskbb} model as the detailed investigation of radiation field for the secondary reflection is beyond the scope of this paper.

%%%%%%%%%%%%%%%%%%%%%%%%%%%%%%%%%%%%%%%%%%%%%%%%%%
\begin{figure}
	\includegraphics[width=\columnwidth, trim={0 30 40 0}]{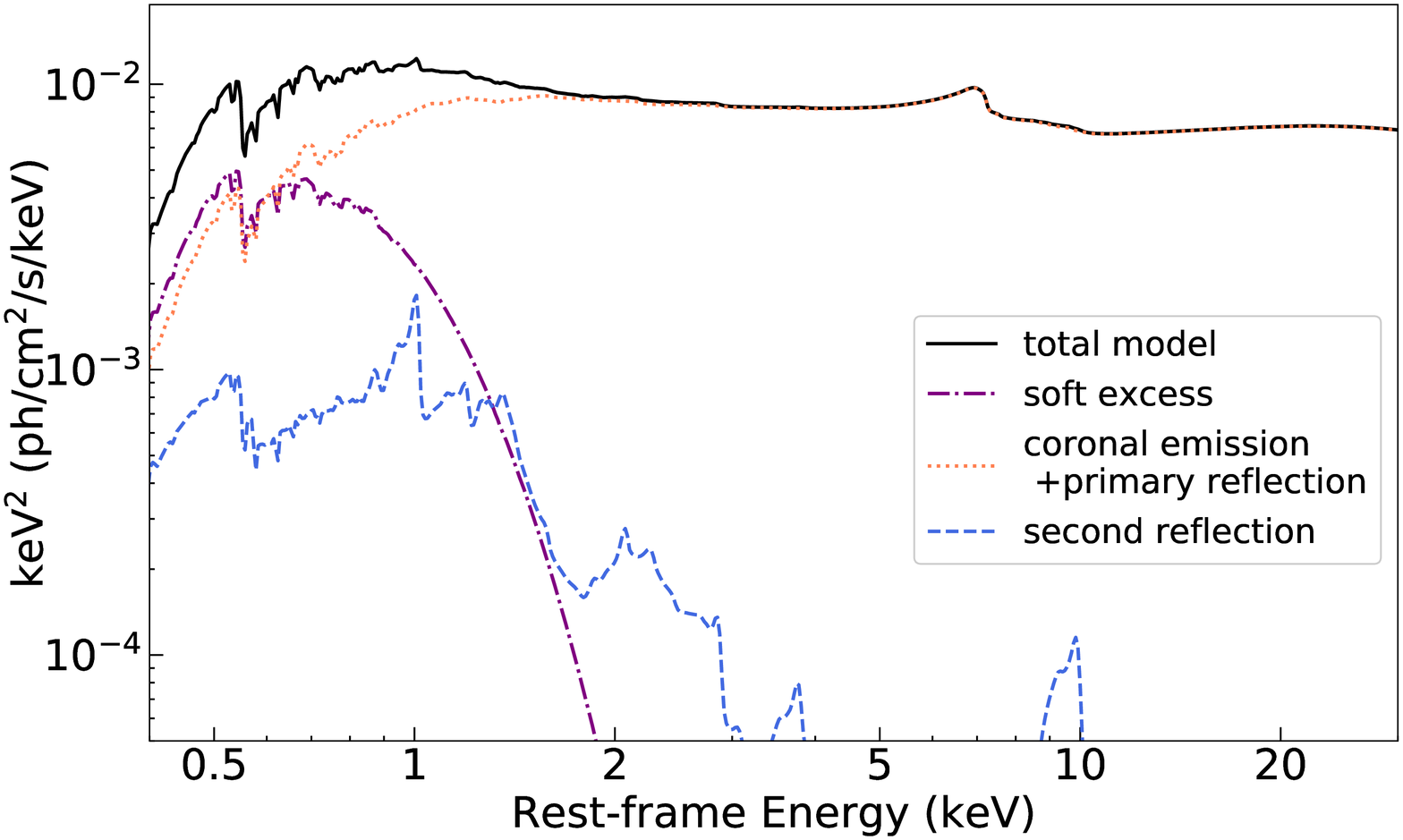}
    \caption{Model components of the best-fit model derived in section~\ref{subsubsec:reflection-emission}. The black solid line is the total contribution of the model components. The purple, orange, and blue dash lines are the soft excess, coronal emission plus the primary relativistic reflection, and the secondary reflection respectively.}
    \label{fig:components}
\end{figure}
%%%%%%%%%%%%%%%%%%%%%%%%%%%%%%%%%%%%%%%%%%%%%%%%%%

Interestingly, a further statistical improvement could be achieved by allowing the inclination angle of the second reflection to vary. We obtain a lower inclination angle ($i=26^{+3}_{-3}$ deg), a smaller inner radius ($R_\mathrm{in}=35^{+17}_{-27}\,R_\mathrm{g}$) and a $\Delta\chi^2/\mathrm{d.o.f.}=7/1$ improvement, corresponding to the significance of $\sim2.7\sigma$. The model components are illustrated in Fig.\ref{fig:components}. The smaller inner radius leads to a skewed broadened line profile due to the relativistic effects rather than a double-peak profile, but the modelling for the 1\,keV emission is even better according to the statistics. As a result, it seems that a strongly blueshifted reflection component with a soft continuum is a physically promising explanation, and the differences caused by the free inclination angle suggest a more complex scenario for the reflection, which is discussed in section~\ref{subsec:emission}. 

%The above scenario naturally drives our investigation of the existence of the possible absorption component with an extremely fast velocity. We re-perform the photoionization model scan with an extended parameter space (up to $z_\mathrm{LOS}=-0.4$) over the time-averaged spectrum with respect to the continuum plus one warm absorber. The result is presented in Fig.\ref{fig:xstar-extension}, showing a secondary peak ($\Delta\chi^2\sim38$) with a high velocity ($>0.3$\,c) and a high ionization state ($\log\xi>3$). However, when we include the third absorber into the spectral fitting, the improvement is only $\Delta\chi^2\sim10$ for the third absorber with $N_\mathrm{H}=5.9^{+0.3}_{-2.6}\times10^{21}\,\mathrm{cm}^{-2}$, $\log(\xi/\mathrm{erg\,cm\,s^{-1})}=4.4^{+0.4}_{-0.1}$ and $z_\mathrm{LOS}=-0.311^{+0.003}_{-0.005}$. Such slight improvement means that there are some degeneracies between the second and third absorber during the fit.  

\section{Discussion}\label{sec:discussion}

\subsection{Multi-phase Absorber}\label{subsec:absorption}
Through the Gaussian line scan and the MC simulations, we detect and confirm several absorption lines at the rest-frame positions of ion transitions and three blueshifted features in the RGS spectrum of 1H 1934, which respectively correspond to a warm absorber and a UFO revealed by the photoionization modelling. The ionization states of two absorbers are similar ($\log\xi\sim1.6\mbox{--}1.7$), while the velocities differ. 

% \citet{2021Laha} has reviewed the possible connections between the warm absorber and the UFO, where the warm absorbers are the cooling and slowing down phases of UFOs.
\citet{2013Tombesi} suggested that the UFO and the warm absorber originate from a single large-scale wind, where the UFO is denser, faster and more ionized than the warm absorber. However, the discovered UFO in 1H 1934 is not highly ionized and even less dense than the warm absorber, inconsistent with this scenario. Alternatively, \citet{2013Pounds} considered another explanation that the warm absorber is produced in the shock where a UFO collides with the surrounding medium. In this case, the weakly ionized UFO discovered in 1H 1934 could be explained by the entrained UFO, which is pushed at a velocity comparable to that of the UFO and retains its ionization state and column density of the surrounding medium. Such entrained UFO (an ultra-fast velocity and low ionization parameter) has also been observed in IRAS 17020+4544 and PG 1114+445, where there is evidence for three kinds of outflows including the warm absorber, entrained-UFO and UFO \citep{2018Sanfrutos,2019Serafinelli}. \yrxu{The entrained UFOs could be the missing link between the high-ionization UFOs and the slow warm absorbers (see more details in \citealt{2021Laha} and references therein).}
%The reason why we do not observe a highly ionized UFO in 1H 1934 probably relates to the relatively low observed angle ($i\sim40\,$deg), thus leading to weak absorption lines. 

Apart from the outflow explanation, an alternative scenario has been proposed that the highly-blueshifted absorption lines seen in the spectrum might be the results of the reflection component passing through a thin, highly-ionized absorbing layer at the surface of the accretion disk \citep[e.g. PG 1211+143 and IRAS 13224-3809;][]{2013Gallo,2020Fabian}. It should be noted that only the part of the absorber in front of the inner reflected emission contributes the absorption. The observed velocity originates from the projected Keplerian motion along our LOS, $v_\mathrm{LOS}=v_\mathrm{abs}\mathrm{sin}\theta\approx v_\mathrm{abs}R_\mathrm{em}/R_\mathrm{abs}$, where $v_\mathrm{abs}$ is the Keplerian velocity of the absorber, $v_\mathrm{abs}=\sqrt{GM/R_\mathrm{abs}}$, and $R_\mathrm{em}$ and $R_\mathrm{abs}$ are the radii of the reflector and absorber respectively. If we assume the reflection occurs at the innermost stable circular orbit (ISCO) of a Schwarzschild black hole ($a_\star=0$), $R_\mathrm{em}=6\,R_\mathrm{g}$, the absorber is thus around $18.6\,R_\mathrm{g}$, where the relativistic beaming effect probably leads to a single blueshifted absorption line. The actual test for this interpretation requires an accurate calculation of the absorption through the disk atmosphere and will be considered in the future paper with more data.

%In this scenario, the blueshift of the absorption is due to the Keplerian motion, and the corresponding radius of the absorber is estimated at $R_\mathrm{K}=GM/v^2\sim166\,R_\mathrm{g}$.

\subsection{Explanations for the Line Emitter}\label{subsec:emission}
The emission features detected in the 1H 1934 spectrum are unknown and, in particular, the 1\,keV emission line has never been well explained in previous work, although a similar broad feature at $\sim$1\,keV was found in 1ES 1927+654 \citep{2021Ricci}. According to our results, we find that the second blueshifted reflection model fits such emission feature better than photo- and collisional-ionization plasma models, provided that the ionizing field is softer ($\Gamma>3.2$) than its primary continuum ($\Gamma\sim2.15$) and the gas is hotter ($\log(\xi/\mathrm{erg\,cm\,s^{-1})}\sim2.5$) than any absorption component ($\log\xi\sim1.6\mbox{--}1.7$). Coupling the inclination of two reflection components ($i\sim39$\,deg) constrains the inner radius of the second reflection to $R_\mathrm{in}\sim70\mbox{--}252\,R_\mathrm{g}$, while leaving them free yields $R_\mathrm{in}\sim8\mbox{--}52\,R_\mathrm{g}$ and a lower inclination angle ($i\sim26\,$deg). The fact that fluorescence provides a better description than recombination of photoionized gas means that the line-emitting gas is likely optically thick, consisting of a layer very close or part of the inner accretion disk. \yrxu{Such plasma is blueshifted as indicated and is outside the line of sight towards the X-ray emitting region (otherwise the soft energy band of the spectra would be highly suppressed).}

% {\bf Collisionally-ionized gas}: A collisional ionization origin is a plausible explanation for the emission features from a statistical point of view. However, from physical aspects, there are several issues about a collision ionization explanation. Collisional ionization is usually expected to result from the shocks between the outflows and surrounding medium or hot confining medium in the inner region. In 1H 1934, the X-ray luminosity ($L_\mathrm{bol}\sim1.725\times10^{44}\,\mathrm{erg}$) is so strong that the photoionization and reflection should be dominant mechanisms. Furthermore, it is unreasonable to only observe the redshifted component ($z_\mathrm{LOS}^\mathrm{CIE}=0.011^{+0.001}_{-0.001}$) in absence of the blueshifted emission. The strong Fe L-complex emission also requires an extraordinary O {\small VIII} flux, which is not observed in the spectrum. Therefore, collisional ionization is disfavored in 1H 1934. 

{\bf Disk atmospheric origin}: The scenario proposed by \citet{2013Gallo} and \citet{2020Fabian} for the blueshifted absorption lines could also be applied to the emission when the disk inclination ($i\sim40$\,deg) is not as large as that of IRAS 13224-3809 ($i\sim70$\,deg), at which the wind absorption may be smaller and more of the inner accretion flow should be observed. In this scenario, the strong blueshift originates from the circulation of the accretion disk, resulting in a velocity shift comparable to the Keplerian velocity expected from a radius close to the $R_\mathrm{in}$. The static disk reflection explanation implies an inclination angle common between the two reflection components. The corresponding inner radius of the second reflection is among $70\mbox{--}252\,R_\mathrm{g}$, which is almost an order of magnitude larger than the Keplerian radius, $R_\mathrm{K}=GM/v^2\sim11\,R_\mathrm{g}$, related to a relativistically corrected velocity ($v\sim-0.293\,c$) from the modelling. Hence, our fits disfavors a purely static disk reflection origin, which requires an inclination angle identical to the primary reflector. 

\begin{figure}
	\includegraphics[width=\columnwidth, trim={50 50  0 0}]{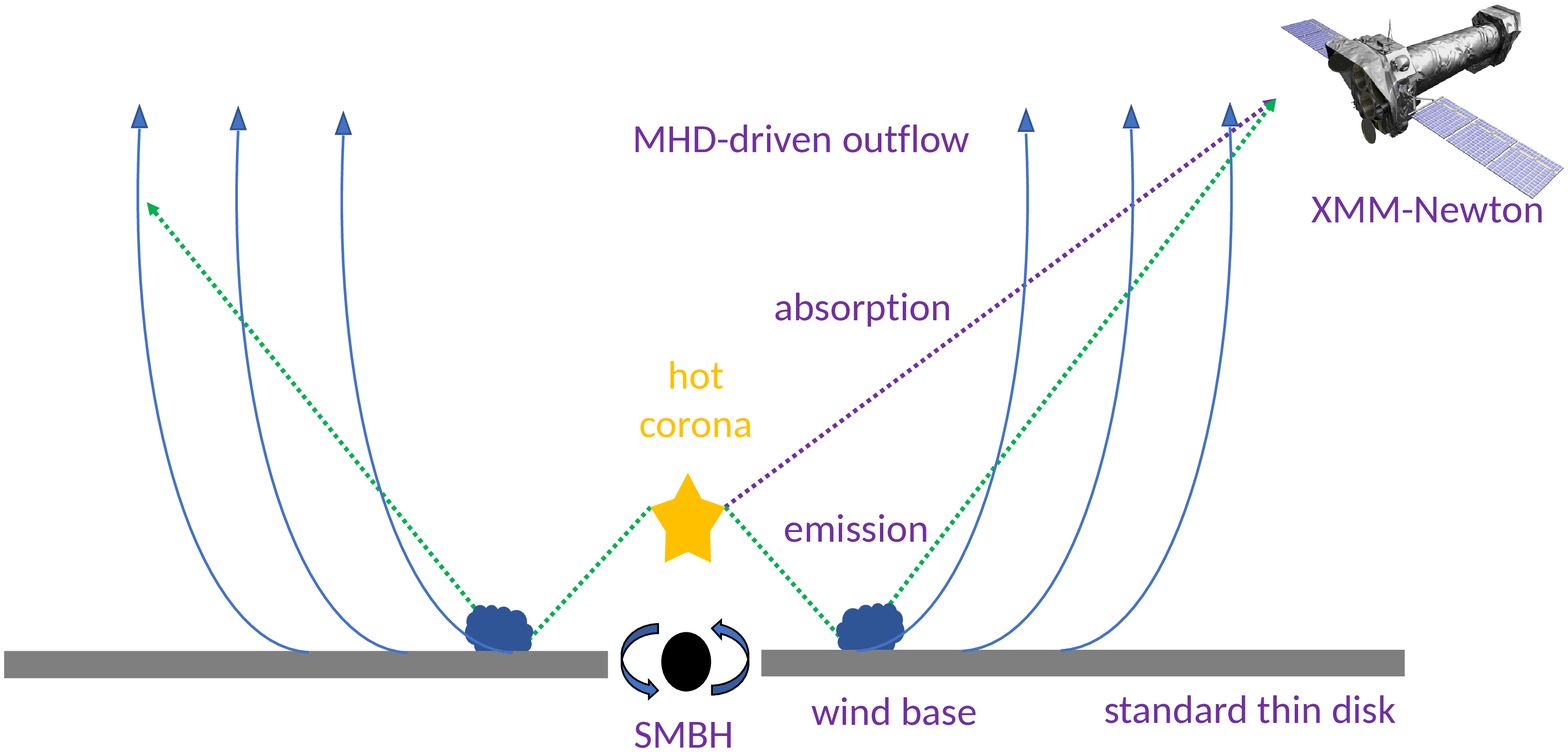}
	\includegraphics[width=\columnwidth, trim={50 50  0 0}]{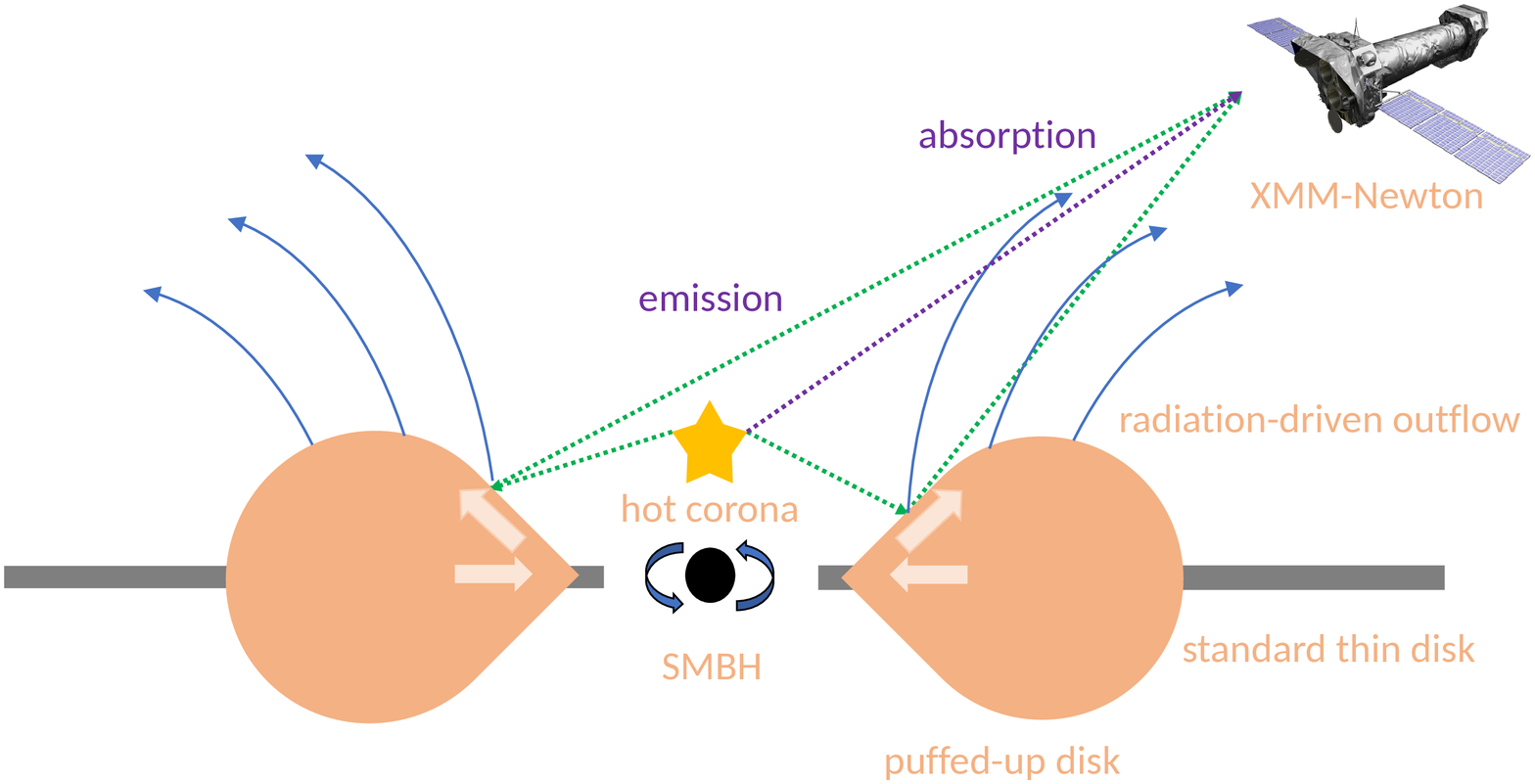}
    \caption{Simplified scheme of the possible explanations for the outflow in 1H 1934-063. The SMBH is surrounded by a standard thin disk ({\it top}) or a thick inner disk ({\it bottom}), and the lamppost geometry is assumed for the hot corona. The absorption lines are observed through the outflow in front of the hot corona. {\it Top panel}: The outflow driven by the magneto-rotational force is launched at a small angle with respect to the disk and then could be quickly lifted to the polar direction. The emission originates from the reflection off the wind base within the disk. {\it Bottom panel}: The radiation pressure in the Eddington-limit system thickens the inner disk and drives the outflow to the equatorial direction. The emission lines comes from the reflection off the inner thick disk.}
    \label{fig:scenarios}
\end{figure}

{\bf Magnetic outflows}: An alternative scenario is that the emission lines are produced by the reflection of the inner accretion flow photons off the base of the magnetically-driven outflow (see the top panel of Fig.\ref{fig:scenarios}). Under this circumstance, the base of the high speed outflow is launched at a small angle with respect to the disk surface and does not obscure the corona but only generates emission lines by reflection (e.g. 1\,keV line). The wind will then be lifted, maximizing the velocity exactly in our LOS, yielding a high blueshift. The absorption features will only be observed when the gas rotates following the magnetic field lines with a velocity vector becoming close to the polar direction. The projection of the velocity in LOS is expected to be smaller than the emission lines, which is consistent with our results. However, it is unclear whether the wind orientation could dramatically change within a small region at a speed of $\sim0.3\,c$, which occurs in a timescale of a few seconds. The requirements of the small region are due to the short recombination time scale of ions \citep[e.g. $\sim1$ sec for O {\small VIII},][]{2020Pintoa} and the fact that outflows will be porous after leaving the base. Moreover, if the velocity of the UFO absorption phase ($\sim0.075\,c$) is indeed due to a projection of the actual velocity, we can estimate the inclination of the LOS with respect to the UFO, $\theta$, by calculating $\theta=\mathrm{arccos}\frac{0.075\,c}{0.293\,c}\sim75\,$deg. The angle between our LOS and the disk axis thus should be $>75\,$deg, implying a Seyfert $1.5\mbox{--}2$ galaxy and more obscuration than what we observed. 
% This angle is also inconsistent with that of the primary reflection. As a result, the magnetically-driven polar outflow is unlikely to explain our results.

%%%%%%%%%%%%%%%%%%%%%%%%%%%%%%%%%%%%%%%%%%%%%%%%%%
%%%%%%%%%%%%%%%%%%%%%%%%%%%%%%%%%%%%%%%%%%%%%%%%%%

{\bf Radiative outflows}: The standard quasar-like equatorial outflow is another possible scenario, where the outflow is launched at an intermediate angle (e.g. close to 40\,deg) with emission lines produced by the reflection of the inner region photons off the base of the wind (see the bottom panel of Fig.\ref{fig:scenarios}). Here the base of wind is the inner thick disk puffed up by the strong radiation field. The primary reflection probably comes from the reprocessing in the inner thin disk. The reflection lines will be observed at a blueshift close to the maximal velocity, while the signatures of the photoionization are hardly detected in the form of either emission or absorption lines because of the over-ionization in the inner region. The UFO moving at $\sim-0.075\,c$ might be the result of the outflow deceleration, or the optically thin outflow from a broad range of radii with a $\sim-0.075\,c$ part making the largest contribution in our LOS. In this case, the wind is inclined with respect to the disk, implying that the inclination angle of the second reflection should be different from the angle between the disk axis and our LOS, compatible with our discovery of a lower inclination angle $i\sim26$\,deg. By assuming the derived velocity as the escape velocity, the launching radius of the outflow would be $R_\mathrm{esc}=2GM/v^2\sim22\,R_\mathrm{g}$, consistent with our fitted $R_\mathrm{in}\sim8\mbox{--}52\,R_\mathrm{g}$. In addition, \citet{2019Thomsen} predicted that the Fe line profile produced from the super-Eddington thick disk is symmetric in shape. The line profile revealed by the line scan (see the 1\,keV emission in the middle panel of Fig.\ref{fig:MC+Gauss}) looks symmetric, supporting the hypothesis of the reflection off a super-Eddington disk. Consequently, the standard quasar-like equatorial outflow could be an explanation for our results to be verified with future observations.

\subsection{Outflow Properties}\label{subsec:properties}
UFOs are launched at small distances from the SMBH but are expected to carry out sufficient kinetic energy ($>0.5\mbox{--}5\%\,L_\mathrm{Edd}$) to affect the evolution of their host galaxies at several orders of magnitude larger scales by quenching or triggering star formation \citep[e.g.][]{2005DiMatteo,2010Hopkins,2013Tombesi,2017Maiolino}. The kinetic energy of the UFO can be expressed as:
\begin{equation}
     L_\mathrm{UFO}=0.5\dot{M}_\mathrm{out}v^2_\mathrm{UFO}=0.5\Omega R^2\rho v_\mathrm{UFO}^3 C_\mathrm{V},
\end{equation}
where $\dot{M}_\mathrm{out}=\Omega R^2\rho v_\mathrm{UFO}C$ is the mass outflow rate, $\Omega$ is the solid angle, $R$ is the distance from the ionizing source, $\rho$ is the outflow density, and $C_\mathrm{V}$ is the volume filling factor (or clumpiness). The density is defined as $\rho=n_\mathrm{H}m_\mathrm{p}\mu$, where $n_\mathrm{H}$ is the number density, $m_\mathrm{p}$ is the proton mass and $\mu=1.2$ is the mean atomic mass assuming Solar abundances. According to the definition of the ionization parameter ($\xi\equiv L_\mathrm{ion}/n_\mathrm{H}R^2$), $n_\mathrm{H}R^2$ could be replaced by the ionization parameter and the ionizing luminosity. Hence, by measuring the ionizing luminosity ($1\mbox{--}1000$\,Rydberg) at $L_\mathrm{ion}\sim1.68\times10^{43}\,\mathrm{erg/s}$, we find:
\begin{equation}
L_\mathrm{UFO}=0.5v_\mathrm{UFO}^3m_\mathrm{p}\mu L_\mathrm{ion}\Omega C/\xi\sim4.77\times10^{45}\Omega C_\mathrm{V}\,\mathrm{erg/s}
\end{equation}
using the UFO results derived from the time-averaged spectrum. If we adopt the conservative value 0.3 for the solid angle from GR-MHD simulations of radiative-driven winds in super-Eddington systems \citep{2013Takeuchi}, and the filling factor $C_\mathrm{V}\sim3\times10^{-4}$ through Eq. 23 in \citet{2018Kobayashi} by assuming that the outflow rate is comparable to the accretion rate, the kinetic energy is $L_\mathrm{UFO}\sim5.2\times10^{42}\,\mathrm{erg/s}\sim3\%L_\mathrm{bol}$ or $1\%L_\mathrm{Edd}$, close to the theoretical criterion to affect the surrounding medium and host galaxy \citep[][]{2005DiMatteo,2010Hopkins}. The detection of reflection off a faster and more ionized phase of the UFO would suggest a stronger feedback.

Furthermore, we attempt to constrain the location of the detected $\sim-0.075\,c$ UFO. If we assume that the UFO velocity is larger than or equal to the escape velocity, the UFO is at least at $R>2GM_\mathrm{BH}/v_\mathrm{UFO}^2\gtrsim 358\,R_\mathrm{g}$. On the other hand, if we hypothesize that the thickness of the UFO is lower than or equal to its maximal distance from the source ($N_\mathrm{H}=C_\mathrm{V}n_\mathrm{H}\Delta R\leq C_\mathrm{V}n_\mathrm{H}R$), the upper limit of the radius is $R\leq C_\mathrm{V}L_\mathrm{ion}/\xi N_\mathrm{H}\sim4.2\times10^{6}\,R_\mathrm{g}$. Hence, the location of that UFO is estimated between $3.6\times10^{2}\mbox{--}4.2\times10^{6}\,R_\mathrm{g}$ and the density of the wind is thus constrained within $n_\mathrm{H}=L_\mathrm{ion}\xi/R^2\sim1.2\times10^{5}\mbox{--}1.7\times10^{13}\,\mathrm{cm}^{-3}$.

\subsection{Comparison with other AGN}\label{subsec:comparison}
Thanks to the high resolution grating spectrometers, the outflow absorption lines and many emission lines have been observed among many other AGN. The properties of the warm absorber in 1H 1934 does not stand out if compared with other AGN. The velocity of the UFO ($0.075\,$c) is at the low end among other AGN, such as Mrk 1044 \citep[0.08\,c,][]{2021Krongold}, PG 1448+273 \citep[0.09\,c,][]{2020Kosec} and IRAS 13224-3809 \citep[0.24\,c,][]{2017Parker}. However, the other parameters of UFO ($\log(N_\mathrm{H}/\mathrm{cm}^{-2})=6.6^{+4.1}_{-2.1}\times10^{19}$ and $\log(\xi/\mathrm{erg\,cm\,s^{-1})}=1.6^{+0.1}_{-0.1}$) are smaller than the typical UFO region, where the ionization parameter and the column density spans from $\log\xi\sim3\mbox{--}6$ and $\log(N_\mathrm{H}/\mathrm{cm}^{-2})\sim22\mbox{--}24$ respectively \citep[e.g.][]{2010Tombesi}. The estimated kinetic energy of UFO is thus weaker but probably still effective enough to affect the host galaxy. For comparison, the similar weak UFO has also been found in other two sources, PG 1114+445 \citep[][]{2019Serafinelli} and IRAS 17020+4544 \citep[][]{2018Sanfrutos}, where the column density of UFO is similar or lower than that of the warm absorber. This can be explained with an entrained UFO, produced by the interaction between the UFO and surrounding materials, fitting in between the UFO and the warm absorber. Futhermore, the location (up to$3.4\times10^{6}\,R_\mathrm{g}$) and the ratio between the kinetic and bolometric energy ($L_\mathrm{UFO}/L_\mathrm{bol}\sim4\%$) of PG 1114+445 are also consistent with those of the UFO in 1H 1934.

%For comparison, a similar weak UFO has also been found in another source, PG 1114+445, where a weakly-ionized ($\log\xi\sim0.5$) UFO with a velocity of $v\sim0.12\,$c  \citep{2019Serafinelli}. The location ($3.4\times10^{6}\,R_\mathrm{g}$) and the ratio between the kinetic and bolometric energy ($L_\mathrm{UFO}/L_\mathrm{bol}\sim4\%$) are also consistent with those of the UFO in 1H 1934. Therefore, the entrained UFO explanation is plausible, although more observations are needed.

The variability of the warm absorber has been observed on a time scale of days or weeks in several Seyfert 1 galaxies, e.g. MCG-6-30-15 \citep[][]{1994Fabian}, Swift J2127.4+5654 \citep[][]{2013Sanfrutos} and Fairall 51 \citep[][]{2015Svoboda}, while the variation only happens in the column density and could be explained by the X-ray eclipse. If the potential variation of the warm absorber in 1H 1934 is true, the change occurs in the ionization state within $\sim1.5$ days due to the variable ionization.

% The soft X-ray emission lines in Seyfert 1 galaxies are more difficult to observe due to the overwhelming contribution of the continuum emission. Nevertheless, with high signal spectra, many emission lines have been observed. The origin of these emission lines is diverse, such as the photoionized emission \citep[e.g. Mrk 335,][]{2019Parker}, scattered emission from the narrow-line region \citep[e.g. Ark 120,][]{2016Reeves}, and the relativistic reflection \citep[e.g. Mrk 110,][]{2021Reeves}. However, 
An origin within relativistic reflection off the base of outflows with a strong blueshift is a rather new topic. Recently, \citet{2016Kara} invoked a strongly blueshifted Fe K reflection line through the X-ray reverberation in a super-Eddington AGN, Swift J1644+57, where the high accretion rate is induced by a tidal disruption event (TDE), in which an AGN destroyed a star and the fallback surpassed Eddington. In 1H 1934, we do not observe a TDE and the same scenario is revealed through the soft X-ray emission rather than Fe K line. We notice that there is a broad emission feature at 1 keV in 1ES 1927+654, which was presumed to be the Ne {\small X} (1.02 keV) or the ionized Fe-L emission from Fe {\small XX-XIV} \citep{2021Ricci}. But another interpretation similar to ours proposed by Masterson et al. (in prep), according to which a high density ($\log n_\mathrm{e}>18\,\mathrm{cm}^{-3}$) and blueshifted ($z\sim-0.3$) reflection model, could reproduce the broad 1 keV line, implying the base of the outflow after the TDE. In our case, we attempt to replace the secondary reflection model with a free-density version and only obtain upper limit of $\log n_\mathrm{e}<16.6\,\mathrm{cm}^{-3}$. It is reasonable since our lines are not as broad as in 1ES 1927+654 and implies the base of the wind instead of the standard disk photosphere, which should be very dense at the inner disk. Ideally, more strongly blueshifted fluorescent lines should be observed to justify this origin, which will be checked with observations. 

\subsection{Future missions}\label{sec:future}
%%%%%%%%%%%%%%%%%%%%%%%%%%%%%%%%%%%%%%%%%%%%%%%%%%
\begin{figure}
	\includegraphics[width=\columnwidth, trim={30 60 50 0}]{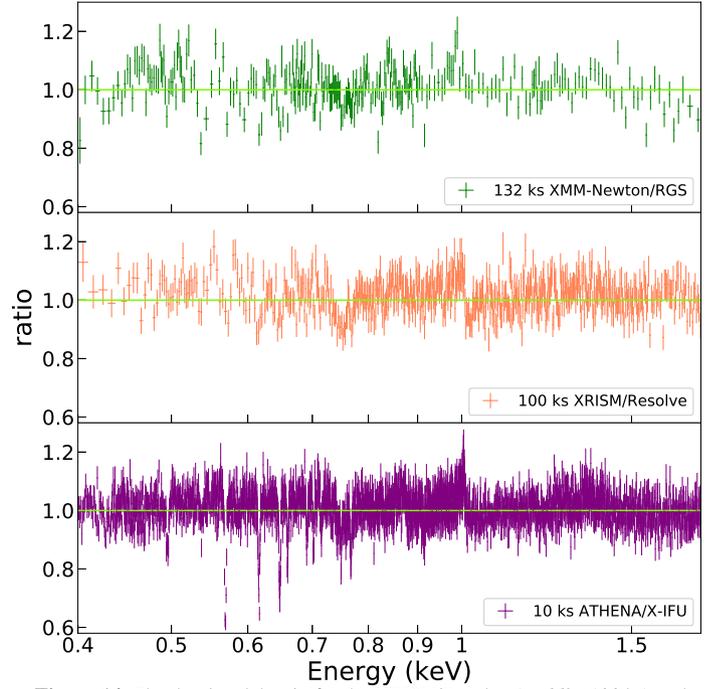}
    \caption{The data/model ratio for the \xrism/Resolve ({\it middle}, 100\,ks) and \athena/X-IFU ({\it bottom}, 10\,ks) spectrum, simulated by the best-fit model obtained in section~\ref{subsubsec:reflection-emission}, with the respect to the baseline continuum model. The ratio for the \xmm/RGS spectrum is shown in the {\it top} panel for comparison.}
    \label{fig:future}
\end{figure}
%%%%%%%%%%%%%%%%%%%%%%%%%%%%%%%%%%%%%%%%%%%%%%%%%%

Future missions with an unprecedented spectral resolution and large collecting areas are expected to improve our understanding of the nature of outflows and decrease the degeneracies among the possible scenarios for the emission. We therefore utilize the \texttt{fakeit} task to simulate data of the X-Ray Imaging and Spectroscopy Mission \citep[\xrism,][]{2018Tashiro} and the Advanced Telescope for High-Energy Astrophysics \citep[\athena,][]{2013Nandra} by using the best-fit model obtained in section~\ref{subsubsec:reflection-emission}. The microcalorimeter (Resolve) onboard \xrism\ provides $5\mbox{--}7$\,eV spectral resolution in the $0.3\mbox{--}12$\,keV bandpass, and the X-ray Integral Field Unit \citep[X-IFU,][]{2018Barret} onboard Athena has an effective energy range of $0.2\mbox{--}12$\,keV with 2.5\,eV spectral resolution up to 7\,keV. We assume an exposure time of 100\,ks for \xrism\ and 10\,ks for \athena. The data/model ratios with respect to the broadband continuum model are shown in the middle and bottom panel of Fig.\ref{fig:future} along with the time-averaged \xmm/RGS spectrum for comparison. The total statistical improvement of the two absorbers and the secondary reflection over the broadband model within $0.4\mbox{--}10.0$\,keV are $\Delta\chi^2\sim250$ and 2990 for \xrism\ and \athena, respectively. Compared with the improvement of the same model in the same energy band of the 140 ks \xmm\ spectrum ($\Delta\chi^2\sim242$), \athena\ reaches over one order of magnitude better significance with one order of magnitude shorter exposure time, while \xrism\ does not provide a much better statistical improvement because most of it comes from the warm absorber. \xrism/Resolve however still has a better spectral resolution than \xmm/RGS above 0.7\,keV, which would be useful to understand the features at 1 keV. 

\begin{figure}
	\includegraphics[width=\columnwidth, trim={30 10 50 0}]{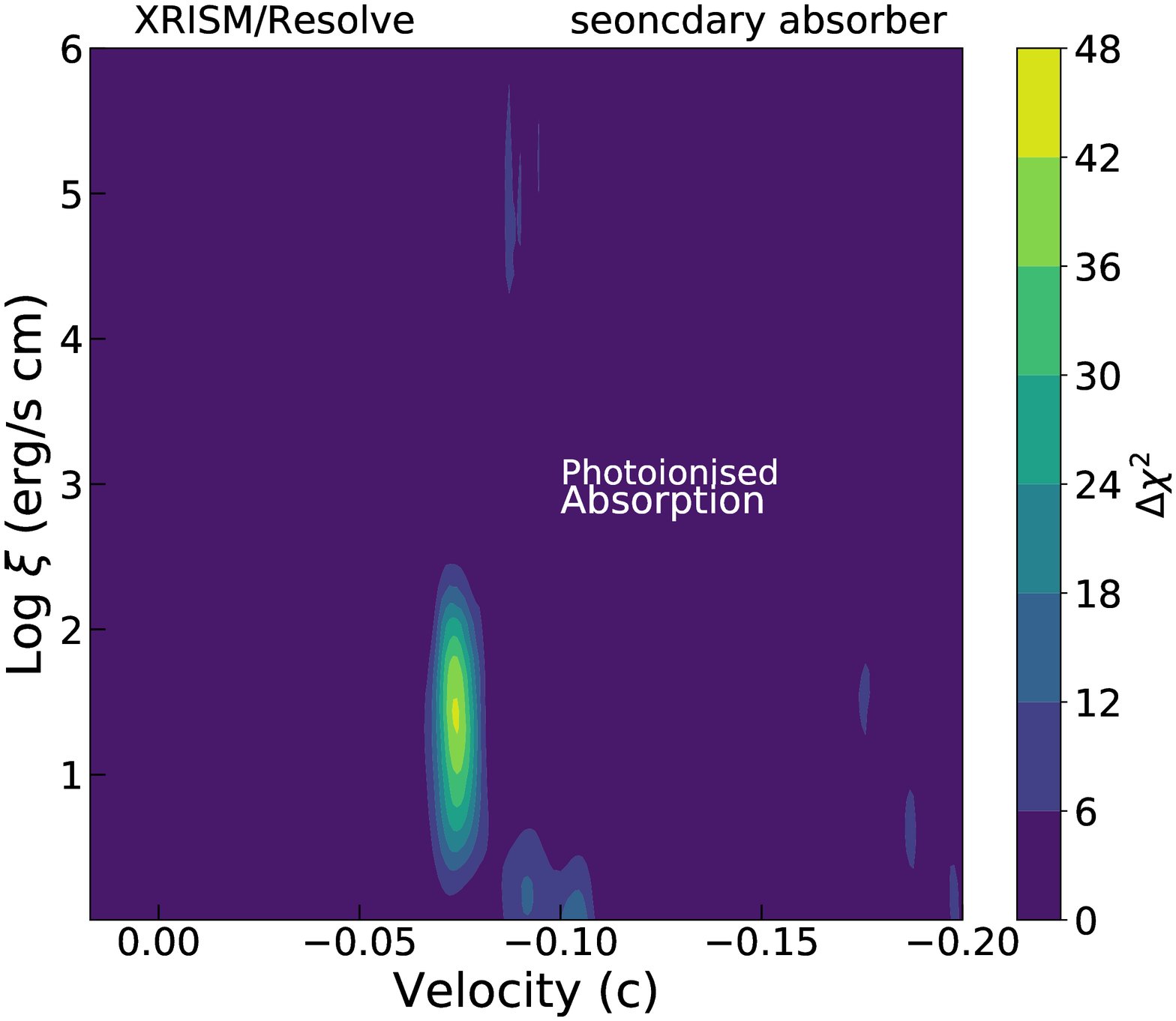}
	\includegraphics[width=\columnwidth, trim={30 10  50 0}]{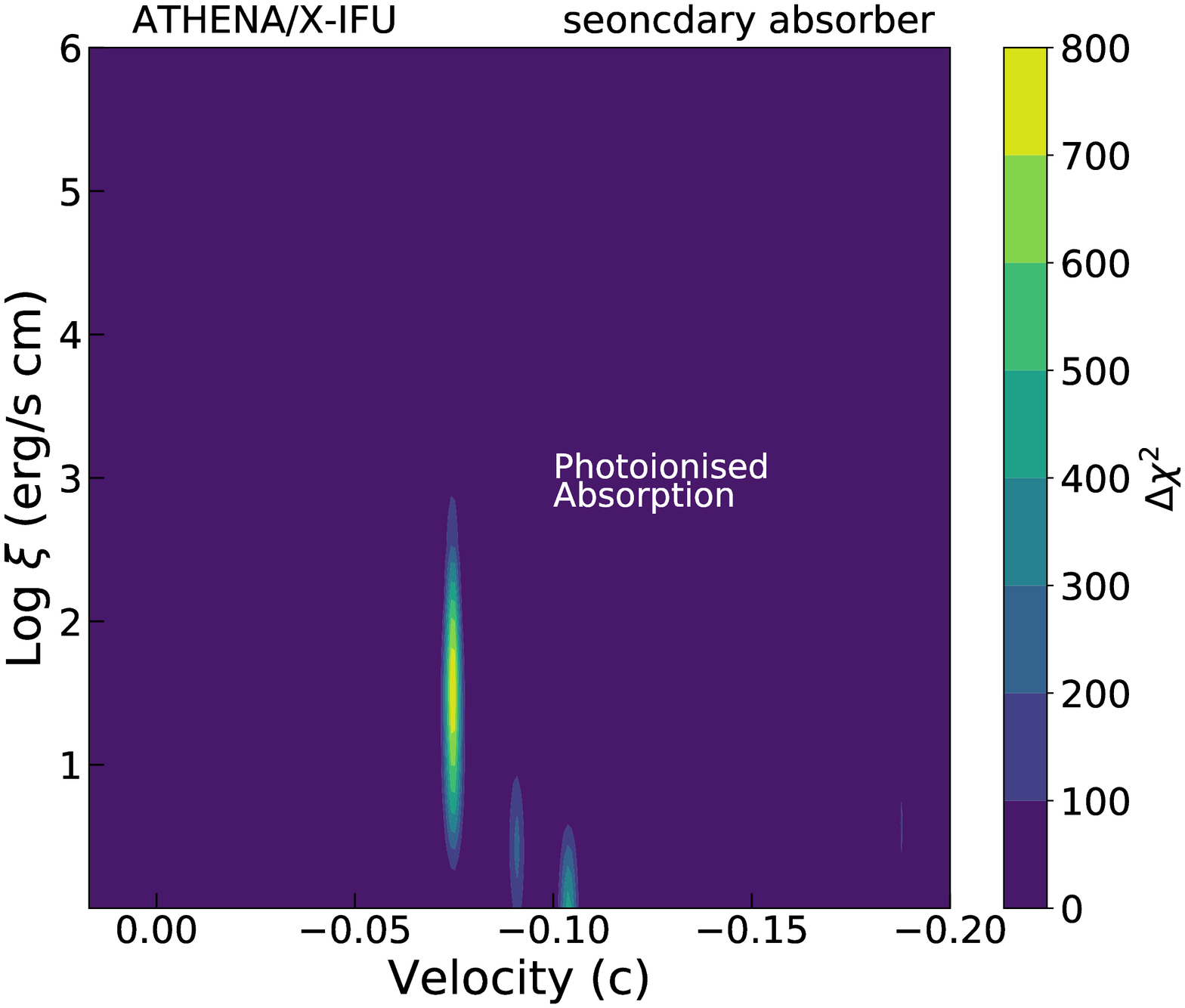}
    \caption{Photoionization absorption model search for the XRISM/Resolve ({\it top}, 100\,ks) and ATHENA/X-IFU ({\it bottom}, 10\,ks) spectrum, simulated with the best-fit model obtained in section~\ref{subsubsec:reflection-emission}, over the continuum model in addition to a warm absorber.}
    \label{fig:future-scan}
\end{figure}

We also perform a photoionization absorption scan on the simulated spectra, where the baseline model is the broadband continuum plus a warm absorber and the second reflection. The results of searching for UFOs are presented in Fig.\ref{fig:future-scan}, where both instruments show a powerful ability to break the degeneracy among different UFO solutions, (see the 'blobs' found in the bottom panel of Fig.\ref{fig:xstar-scan}). The result of \xrism\ presents a comparable detection significance with \xmm\ and that of \athena\ gives a remarkably better detection significance with a much shorter time scale. Therefore, the future missions are promisingly able to deepen our understanding of AGN outflows as well as the nature of the soft X-ray lines.

%%%%%%%%%%%%%%%%%%%%%%%%%%%%%%%%%%%%%%%%%%%%%%%%%%
%%%%%%%%%%%%%%%%%%%%%%%%%%%%%%%%%%%%%%%%%%%%%%%%%%

\section{Conclusions}\label{sec:conclusion}
In this work, we perform a variability analysis, and a time- and flux-resolved X-ray spectroscopy on a joint \xmm\ and \nustar\ observation of NLS1 1H 1934-063 in 2015 to investigate the nature of the soft X-ray features. We find some absorption features are close to their rest frame and come from a distant warm absorber, which seems to weakly vary within the exposure time. Some absorption features are consistent with a UFO ($v_\mathrm{UFO}\sim-0.075\,c$), which might be an entrained UFO as a result of the interaction between the UFO and surrounding medium. The detected emission lines do not matched any known rest-frame ion transitions. A secondary blueshifted ($z\sim-0.3\,c$) reflection model fits such emission features better than photo- or collisional-ionization plasma models. We explain this with the reprocessing of inner accretion flow photons off the base of an equatorial wind, which could be the link between reflection and ejection in high-accretion AGN. Future observations will be useful to test this scenario.

\section*{Acknowledgements}
E.K. acknowledges support from NASA ADAP grant No. 80NSSC17K0515. J.A.G. acknowledges support from NASA ATP grant 80NSSC20K0540 and from the Alexander von Humboldt Foundation. W.N.A. is supported by an ESA research fellowship.

%The Acknowledgements section is not numbered. Here you can thank helpful colleagues, acknowledge funding agencies, telescopes and facilities used etc. Try to keep it short.

%%%%%%%%%%%%%%%%%%%%%%%%%%%%%%%%%%%%%%%%%%%%%%%%%%
\section*{Data Availability}
The \xmm\ and \nustar\ data underlying this article are available in ESA's XMM-Newton Science Archive (https://www.cosmos.esa.int/web/xmm-newton/xsa) and HEASARC NuSTAR Archive respectively. 

%The inclusion of a Data Availability Statement is a requirement for articles published in MNRAS. Data Availability Statements provide a standardised format for readers to understand the availability of data underlying the research results described in the article. The statement may refer to original data generated in the course of the study or to third-party data analysed in the article. The statement should describe and provide means of access, where possible, by linking to the data or providing the required accession numbers for the relevant databases or DOIs.



\bibliographystyle{mnras}
\bibliography{ref} % if your bibtex file is called example.bib

%\begin{thebibliography}{99}
%\end{thebibliography}

%%%%%%%%%%%%%%%%%%%%%%%%%%%%%%%%%%%%%%%%%%%%%%%%%%

%%%%%%%%%%%%%%%%% APPENDICES %%%%%%%%%%%%%%%%%%%%%

\appendix
\section{Systematic Effect}\label{subsec:system}
Flux-resolved spectroscopy could introduce some issues as it is based on the assumption that the dominant emission/absorption mechanisms do not change during the observation. However should the large dip event be produced by a different process than those responsible for the low flux epochs throughout the observation, it might affect our results. In addition, when we merge the spectra of the same fluxes but different epochs, the time variability could smear or broaden some spectral features. Therefore, we also performed a time-resolved spectroscopy to verify our results. 

%%%%%%%%%%%%%%%%%%%%%%%%%%%%%%%%%%%%%%%%%%%%%%%%%%
\begin{figure}
	\includegraphics[width=\columnwidth, trim={50 30  50 0}]{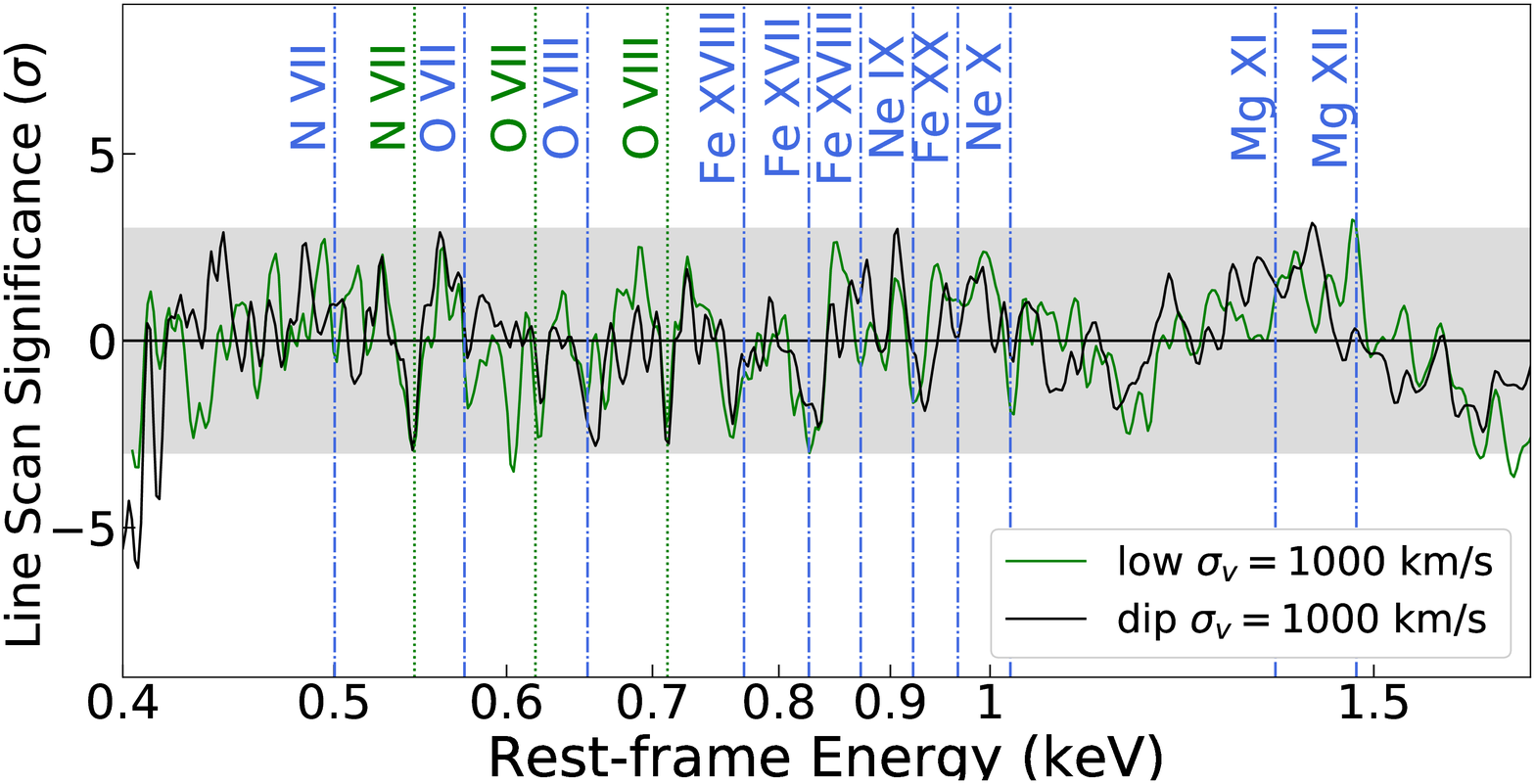}
    \caption{The line search result of the dip spectrum over the RGS band with a velocity line width of 1000\,km/s. The result of the low-flux spectrum is also overlapped for comparison.}
    \label{fig:dip}
\end{figure}
%%%%%%%%%%%%%%%%%%%%%%%%%%%%%%%%%%%%%%%%%%%%%%%%%%

As shown in the left panel of Fig.\ref{fig:lc_histogram}, we also extract the time-resolved \xmm\ and \nustar\ spectra taken during the flux dip, displayed with the vertical grey region. A Gaussian line scan over the RGS band is launched adopting the broadband continuum model used in section~\ref{subsec:line-scan}. The result is presented in Fig.\ref{fig:dip}, where the result for the low-flux spectra is also illustrated for comparison. We found that the shapes of the scan over the flux-resolved and time-resolved spectra are generally compatible, with both the warm absorber and UFO absorption lines being consistent. This confirms that any systematics driven by our choice of performing a flux- rather than time-resolved spectroscopy should not affect our conclusions. 

% % \section{Some extra materials}
% %%%%%%%%%%%%%%%%%%%%%%%%%%%%%%%%%%%%%%%%%%%%%%%%%%
% \begin{figure}
% 	\includegraphics[width=\columnwidth, trim={50 100 30 0}]{PN_PCA_components.eps}
%     \caption{Component spectra for the first three PCs from the analysis of the \xmm\ EPIC-pn data of 1H 1934. Energies are in the rest frame. The first PC indicates a variable power law. The second PC presents the pivoting effect of the primary continuum and the third PC component is contaminated by noise.}
%     \label{fig:PN-PCA}
% \end{figure}
% %%%%%%%%%%%%%%%%%%%%%%%%%%%%%%%%%%%%%%%%%%%%%%%%%%
\section{Cross-correlation Monte Carlo Simulation}\label{sec:CC-MC}
\begin{figure}
	\includegraphics[width=\columnwidth, trim={50 100 50 0}]{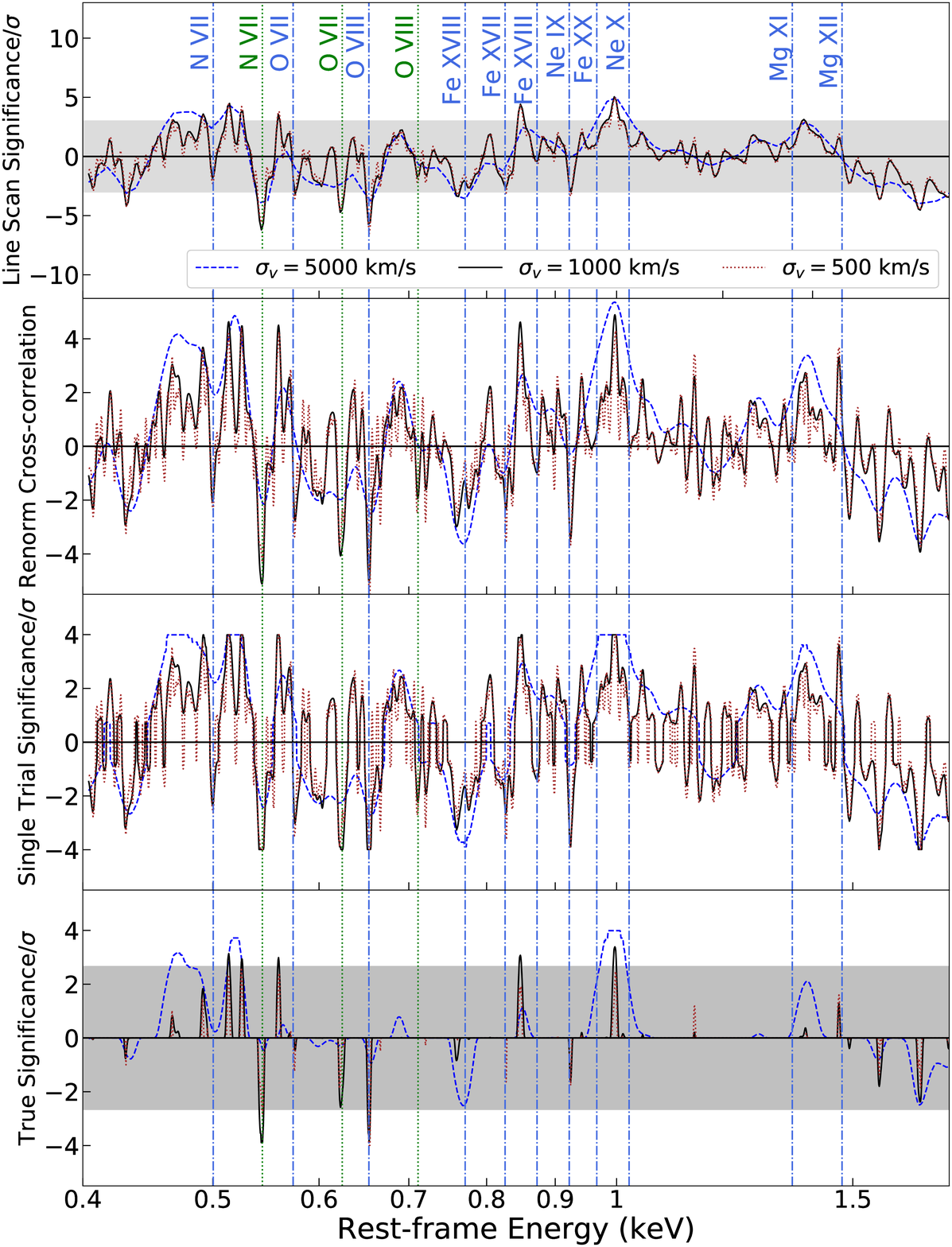}
    \caption{The zoom-in Gaussian line result (same as the \yrxu{top} panel of Fig.\ref{fig:MC+Gauss} for comparison), renormalized cross-correlation, single trial significance and the true significance (same as the bottom panel of Fig.\ref{fig:MC+Gauss}) for the time-averaged RGS spectra (from top to bottom panel). The maximum and minimum of the single trial and true significance are manually set at $\pm4\sigma$ as we only search on 10000 simulated spectra. The single trial significance obtained from MC simulation, as expected, consistent with the Gaussian line scan.}
    \label{fig:RGS-simulation}
\end{figure}

Our procedures after modelling the continuum are briefly summarized below: 1) we generate and save the RGS residual spectrum of the source, where the Y-axis is in a unit of Photon/s/keV; 2) we use the {\small fakeit} tool in XSPEC to simulate 10000 spectra based on the best-fit continuum model with the same exposure time as the real RGS spectrum and record the residual spectra as well; 3) we generate a set of Gaussian line models, whose centroid energy spans over the RGS band with a logarithmic grid of 2000 points, line width ranges from 0 to 5000\,km/s, normalization is fixed at unity, and the energy bins are the same as those of the source spectrum; 4) we cross-correlate the real residual spectrum with all the generated Gaussian line models in a form of
$$C=\sum\limits^N_{i=1}x_iy_i,$$
where $x$ and $y$ are respectively the arrays of the real residual spectrum and the Gaussian line model at a predefined width and centroid energy; 5) we cross-correlate the simulated datasets with all Gaussian line models, similarly done with the real data; 6) we renormalize the raw cross-correlations for both real and simulated data; 7) we calculate the p-value of each bin in the real dataset, i.e. the fraction of the simulated datasets that show cross-correlation stronger than the real data, which gives the single trial significance; 8) we obtain the true p-value of each cross-correlation in the real data by calculating the fraction of the correlation and anti-correlation from each simulated dataset (i.e. for emission and absorption lines found anywhere within RGS band) larger than the real one. Hence, the true p-value takes the look-elsewhere effect into account. The results of the MC simulations with 500, 1000, and 5000 km/s are shown in detail in Fig.\ref{fig:RGS-simulation}. For more details, we refer to Appendix B in \citet{2021Kosec}.

%%%%%%%%%%%%%%%%%%%%%%%%%%%%%%%%%%%%%%%%%%%%%%%%%%
%%%%%%%%%%%%%%%%%%%%%%%%%%%%%%%%%%%%%%%%%%%%%%%%%%

% %%%%%%%%%%%%%%%%%%%%%%%%%%%%%%%%%%%%%%%%%%%%%%%%%%
% \begin{figure}
% 	\includegraphics[width=\columnwidth, trim={50 50  50 0}]{double_absorption_100_avg_extension.eps}
%     \caption{The same scan as the bottom panel of Fig.\ref{fig:xstar-scan} with an extended searched parameter space, up to $z_\mathrm{LOS}=-0.4$, showing a secondary highly-ionized and fast solution ($\log\xi>3$ and $v>0.3$\,c).}
%     \label{fig:xstar-extension}
% \end{figure}
% %%%%%%%%%%%%%%%%%%%%%%%%%%%%%%%%%%%%%%%%%%%%%%%%%%

% Don't change these lines
\bsp	% typesetting comment
\label{lastpage}
\end{document}